\documentclass[twocolumn,pra,aps,superscriptaddress,showpacs]{revtex4-1}

 \usepackage{mathptmx}
\usepackage{dcolumn}
\usepackage{amsmath,amssymb,amsfonts}
\usepackage{dsfont}
\usepackage{bm}
\usepackage{color}
\usepackage{latexsym}
\usepackage{epstopdf}
\usepackage[english]{babel}
\usepackage{enumerate}

\usepackage[pdftex]{graphicx}

\usepackage{natbib}
\usepackage{epstopdf}
\usepackage{appendix}

\definecolor{mygrey}{gray}{0.35}
\definecolor{myblue}{rgb}{0.2,0.2,0.8}
\definecolor{myzard}{cmyk}{0,0,0.05,0}
\definecolor{mywhite}{rgb}{1,1,1}
\definecolor{mywhite}{rgb}{1,1,1}
\definecolor{myred}{rgb}{1,0.,0.3}

\usepackage[colorlinks=true,citecolor=myblue,linkcolor=myred]{hyperref}

\newcommand{\bra}[1]{\left\langle #1\right|}
\newcommand{\ket}[1]{\left| #1\right\rangle}

\newcommand{\mean}[1]{\langle #1\rangle}

\newcommand\nn{\mathbf{n}}
\newcommand\mm{\mathbf{m}}
\newcommand\kk{\mathbf{k}}
\newcommand\qq{\mathbf{q}}

\newcommand\vv{\mathbf{v}}

\newcommand\KK{K}

\newcommand\intt{\mathrm{int}}

\newcommand\BS{\mathrm{BS}}

\begin{document}

\title{Markovian and Non-Markovian Dynamics of Quantum Emitters coupled to Two-dimensional Structured Reservoirs}
 \author{A. Gonz\'{a}lez-Tudela}
 \email{alejandro.gonzalez-tudela@mpq.mpg.de}
 \affiliation{Max-Planck-Institut f\"{u}r Quantenoptik Hans-Kopfermann-Str. 1. 85748 Garching, Germany }
 \author{J. I. Cirac}
 \affiliation{Max-Planck-Institut f\"{u}r Quantenoptik Hans-Kopfermann-Str. 1. 85748 Garching, Germany }

\begin{abstract}

The interaction of quantum emitters with structured baths modifies both their individual and collective dynamics. In Ref.~\cite{gonzaleztudela17a} we show how exotic quantum dynamics emerge when QEs are spectrally tuned around the middle of the band of a two-dimensional structured reservoir, where we predict the failure of perturbative treatments, anisotropic non-markovian interactions and novel super and subradiant behaviour. In this work, we provide further analysis of that situation, together with a complete analysis for the quantum emitter dynamics in spectral regions different from the center of the band.
\end{abstract}

\maketitle

\section{Introduction \label{sec:intro}}

Even in perfect isolation from other systems, an optical quantum emitter (QE) interacts unavoidably with the bath of vacuum photons, which renormalizes its energy and gives it a finite lifetime~\cite{cohenbook92a}. Interestingly, when several QEs are present they can exchange interactions through the photon bath. This generates both (coherent) interactions and collective decay between them which quickly decay with the distance between QEs in the near field, due to energy spread into the whole solid angle \cite{lehmberg70a,lehmberg70b}. Coherent interactions are instrumental for designing quantum gates or spin-exchange interactions for quantum simulation, whereas collective decay renormalizes the lifetime of individual atoms generating super and subradiant states \cite{dicke54a} decaying faster/slower than they would do independently. However, in structureless 3D baths both mechanisms compete \cite{gross82a} and make it difficult to observe these effects in a clean way.

Since Purcell predicted the modification of QE decay rates inside cavities~\cite{purcell46a}, the possibility of modifying QE dynamics through spectral shaping of the vacuum modes has attracted a lot of interest (see ~\cite{nakazato96a,lambropoulos00a,woldeyohannes03a,giraldi11a,breuer16a,devega17a} and references therein for several reviews on the subject).
Recent advances in QE-nanophotonics integration~\cite{vetsch10a,huck11a,hausmann12a,laucht12a,thompson13a,goban13a,beguin14a,lodahl15a,bermudez15a,sipahigi16a,corzo16a,sorensen16a,sipahigi16a,solano17a} and the possibility of mimicking such QED scenarios in circuit QED \cite{houck12a,astafiev10a,hoi11a,vanloo13a,liu17a} or cold atoms \cite{devega08a,navarretebenlloch11a} has increased the interest in studying systems where the baths not only have an spectral structure but are also confined to reduced dimensionalities. This interplay between the structure and reduced dimensionality results in qualitatively new physics.

For example, in the case of one-dimensional (1D) baths the reduced dimensionality induces infinite range collective interactions between the QEs in which the dipole-dipole couplings can be made zero, while keeping the collective decay maximal~\cite{kien05a,dzsotjan10a,gonzaleztudela11a,chang12a,shahmoon13a}. This leads to perfect super/subradiance~\cite{goban15a,corzo16a,sorensen16a,sipahigi16a,solano17a}, which can be exploited for entanglement generation, self-organization of atoms or multiphoton generation among others~\cite{gonzaleztudela11a,gonzaleztudela13a,facchi16a,chang13a,chang12a,corzo16a,gonzaleztudela15a,gonzaleztudela16a}.

Apart from the effects introduced by the reduced dimensionality, further flexibility is obtained with \emph{structured} reservoirs, as they naturally appear in photonic crystals \cite{joannopoulos_book95a} or atoms in optical lattices~\cite{devega08a,navarretebenlloch11a}. By structured we mean possesing a periodic structure that gives rise to the existence of bands in the dispersion relation of the propagating modes. For example, placing the QE energies within the bandgap leads to the localization of photons around them~\cite{bykov75a,john90a,kurizki90a,longo10a,shi16a,calajo16a,sanchezburillo16a}, which can mediate long-range purely coherent interactions between the QEs~\cite{douglas15a,gonzaleztudela15c}. Last, but not least, the interplay of the confinement of the fields and the polarization of light, allows one to control the directionality of the emission~\cite{mitsch14a,sollner15a}, leading to the so-called \emph{chiral quantum optics}~\cite{lodahl16a}. This 1D directional emission has been shown to 
generate novel many-body entangled steady states \cite{ramos14a}, non-reciprocal photon transport~\cite{sayrin15a} and more efficient implementations for quantum networks~\cite{pichler15a,mahmoodian16a} and computation \cite{pichler17a}.

From all those examples, it is clear that quantum optics with low dimensional structured baths leads to new type of interactions, which afterwards can be harnessed for applications impossible to obtain otherwise. In this work, we focus on the coupling of QEs to two-dimensional baths (2D). In particular, we consider the \emph{structured} 2D reservoir of bosonic modes with square symmetry that we studied in the main manuscript \cite{gonzaleztudela17a}, and focus on the situation where the QE transition frequency lies within the band. Apart from the long range character of the interactions expected from the reduced dimensionality, we analyze several effects such as non-perturbative relaxation of a single QE which is accompanied by directional emission into the bath, as also predicted for a classical source of light~\cite{mekis99a} and sound~\cite{langley96a}. The interplay between the directionality, position and relative phases between QEs leads to super/subradiant behaviour for two QEs, This phenomenon has 
been discussed for a similar model in Ref.~\cite{galve17a} within a Born-Markov master equation 
description. Using an exact treatment, we find that perfect subradiance/superradiance is not possible for two QEs, i.e., there exist no states with a finite excitation in the limit $t\rightarrow \infty$. We give an intuitive explanation of this fact in terms of the interference of the light emerging from both emitters, which can only be destructive in one of the propagation directions. However, for four QEs we are able to find a situation where all the emitted light interferes, and thus complete subradiance could be observed. 

The outline of this manuscript reads as follows: first, in Section~\ref{sec:system} we introduce the system of interest and in Section \ref{sec:theory} we explain both the theoretical and numerical tools that we use to characterize it respectively. Before moving to the core of the results of the paper, we review the results obtained already for the 1D scenario in Section \ref{sec:1D}, that will help us to emphasize what is different for our 2D situation. Then, in Sections~\ref{sec:single},\ref{sec:two} and \ref{sec:many}, we study the situation where one, two and many QEs are interacting through the 2D environment. In Section~\ref{sec:experimental} we make a brief discussion on the experimental feasibility in two platforms and analyze the impact of some experimental limitations of the phenomena predicted. Finally, we summarize the main results and point to future directions of work in Section \ref{sec:conclu}.

\section{System \label{sec:system}}

\begin{figure}
\centering
\includegraphics[width=0.8\linewidth]{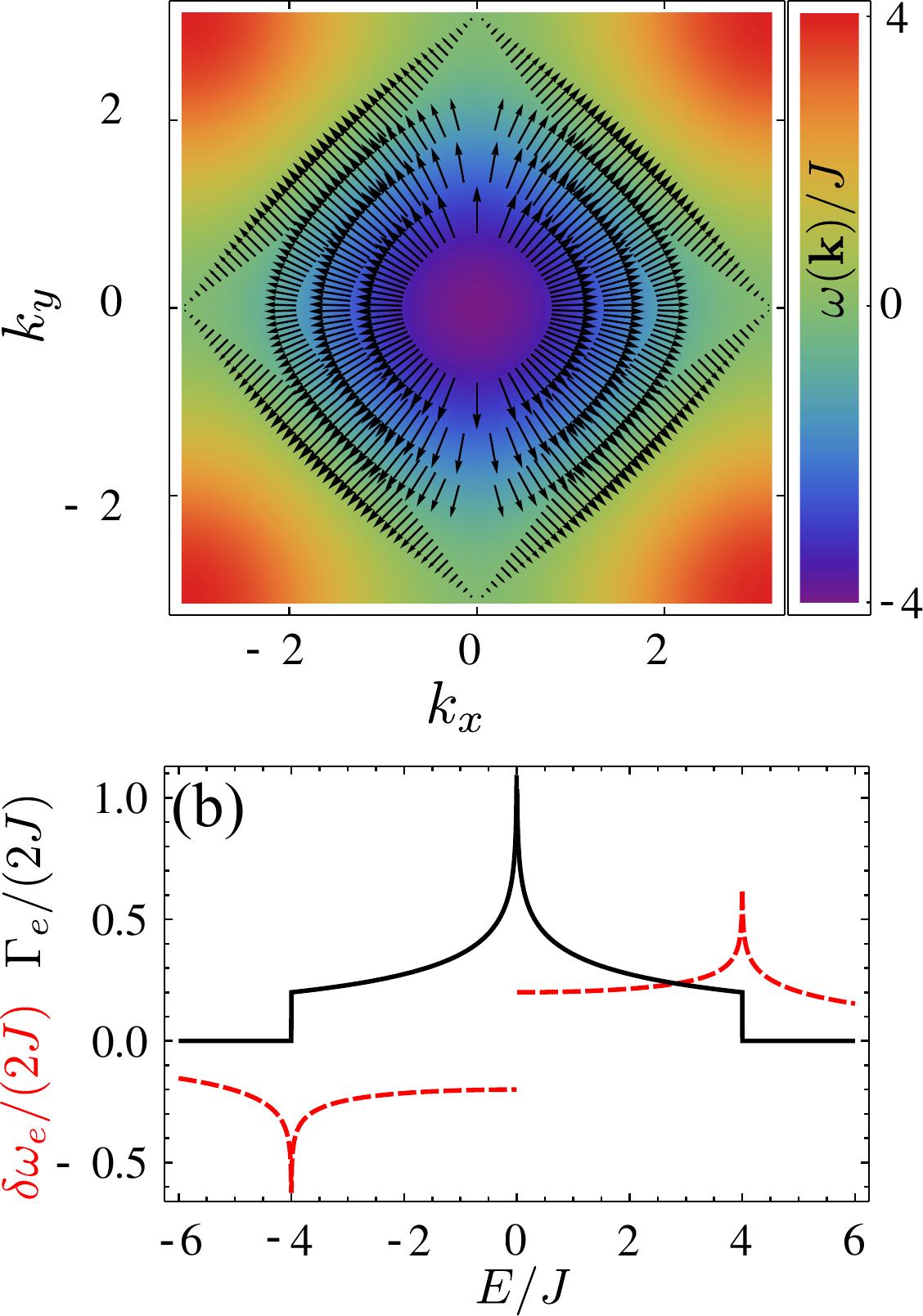}
\caption{(a) Contour plot of $\omega(\kk)$ for the square lattice tight-binding model with the vector fields of the group velocities for $E/J=(-3,-2,-1,0)$. (b) Real $\delta\omega_e$ (dotted red) and imaginary $\Gamma_e$ [proportional to the density of states $D(E)$] (solid black) part of the single QE self-energy $\Sigma_{e}(E)$ for $g=J$. }
\label{fig1L}
\end{figure}

In this work we are interested in a bath which has a 2D square lattice structure. Given the variety of experimental systems that are available nowadays to obtain the QED-like Hamiltonians~\cite{vetsch10a,huck11a,hausmann12a,laucht12a,thompson13a,goban13a,beguin14a,lodahl15a,bermudez15a,sipahigi16a,houck12a,astafiev10a,hoi11a,vanloo13a,liu17a,devega08a,navarretebenlloch11a}, we consider a simplified model to describe the reservoir that can capture the most important features of the system, in the same spirit than using a bosonic tight-binding model for 1D systems. In particular, we describe our 2D bath as a set of $N\times N$ bosonic modes, with annihilation operators $a_\nn$, disposed in a square lattice, with energies $\omega_a$, position described by two integer indices $\nn=(n_x,n_y)$ (as we take the distance $d\equiv 1$ as the length unit) and nearest neighbor coupling $J$. Thus, its free Hamiltonian reads (we use $\hbar=1$ along the manuscript):
\begin{equation}
 \label{eql:HB}
 H_B=-J\sum_{\mean{\nn,\mm}} \left(a_\nn^\dagger a_\mm+\mathrm{h.c.}\right)\,.
\end{equation}

Notice that we have written the Hamiltonian $H_B$ in a frame rotating with $\omega_a$, that we use to write all the Hamiltonians along this manuscript. The bath Hamiltonian can be diagonalized by imposing periodic boundary conditions and introducing the operators $a_\nn=\frac{1}{N}\sum_{\kk} a_\kk e^{i\kk\cdot \nn}$, where $\kk=(k_x,k_y)$, with $k_i\in \frac{2\pi}{N}(-\frac{N}{2},\dots,\frac{N}{2}-1)$. In that basis, $H_B=\sum_{\kk}\omega(\kk)a^\dagger_\kk a_\kk$, where $\omega(\kk)=-2 J\left[\cos(k_x)+\cos(k_y)\right]$ that we plot in Fig.~\ref{fig1L}. Despite being a very simplified model, it captures several interesting characteristics of 2D reservoirs:
\begin{itemize}
 \item For the single excitation subspace it captures the effect of having a band with a finite bandwidth, i.e., $\omega(\kk)\in [-4 J,4J]$. Therefore, it allows one to explore the situation when the QE transition frequency lies in or out of the band.
 
 \item Close to the band-edges $\mp 4J$, the energy dispersion can be shown to be isotropic,  therefore capturing the physics of other simplified models used in the literature \cite{gonzaleztudela15c}. For example, for the lower edge, i.e., $\omega(\kk)\approx - 4J+ J |\kk|^2$ for $|\kk|^2\ll 1$
 
 \item At the X,Y points the energy dispersion displays a saddle point, e.g., at $\kk_X=(0,\pi)$, $\omega(\kk_X+\qq)\propto q_y^2-q_x^2$, for $q_{x,y}\ll 1$. This saddle point gives rise to a divergent density of states in the thermodynamic limit, defined as:
 \begin{equation}
 D(E)=\frac{1}{(2\pi)^2}\iint d\kk\delta\left[E-\omega(\kk)\right]\,.
  \end{equation}
  
 These saddle points are ubiquitous in 2D systems, also with other geometries, because minima have to be connected within different Brillouin zones as it happens, e.g., in real materials~\cite{mekis99a,langley96a}. Thus, we expect the dynamics we find associated to such divergences to be observed in many different systems.
 
 \item As the energy moves close to the middle of the band the $\kk$-line of equal frequencies $E=\omega(\kk)$ gets more anisotropic as depicted by plotting the vector field associated to the group velocity $\vv_g(\kk)=2J(\sin(k_x),\sin(k_y))$ at the $\kk$'s satisfying $\omega(\kk)=E$. At exactly the middle point, i.e., $E=0$, there are $\kk$ points in which the group velocity is zero, $|\vv_g(0,\pm \pi)|=|\vv_g(\pm\pi,0)|=0$ and others where it is maximal. Note that all this can be associated to the appearance of the saddle points mentioned in the previous item~\cite{mekis99a,langley96a}.
 
\end{itemize}

We describe our QEs as two-level systems $\{\ket{g}_j,\ket{e}_j\}$ with transition frequency, $\omega_e$. Thus, their bare Hamiltonian in the frame rotating with $\omega_a$ is given by $H_S=\Delta\sum_{j=1}^{N_e} \sigma_{ee}^j$, where $\Delta=\omega_e-\omega_a$ and where we use the notation $\sigma_{\alpha\beta}^j=\ket{\alpha}_j\bra{\beta}$ for the spin operators. The QEs have a transition $\ket{e}\leftrightarrow\ket{g}$ which couples to the bath modes locally, such that their Hamiltonian is given by:
\begin{equation}
\label{eqL:Hintt}
 H_\mathrm{int}=g\sum_{j=1}^{N_e} \left(a_{\nn_j} \sigma_{eg}^j+\mathrm{h.c.}\right)\,.
\end{equation}
where $\nn_j$ denotes the position of the bath mode which interacts with the $j$-th QE. Notice that, for typical electromagnetic dipole coupling, we have performed the rotating wave approximation to neglect the counter-rotating terms, which requires that $\omega_a,\omega_e\gg g$; that is a safe assumption in the optical regime. As a consequence, the total Hamiltonian $H=H_S+H_B+H_\intt$ conserves the number of excitations $N_\mathrm{exc}=\sum_{j}\sigma_{ee}^j+\sum_{\nn}a_\nn^\dagger a_\nn$, which allows us to diagonalize in each excitation subspace. We are also considering the coupling to a single polarization of light, such that one can assume a scalar description of the bosonic field, and also to a single bosonic band. Considering the coupling to several bands may also give rise to interesting phenomena as shown in Ref.~\cite{gonzaleztudela17d}, but this is beyond the scope of this paper.

Despite its simplicity this model can be implemented in several platforms. For example, using alkaline-earth atoms in state dependent optical lattices~\cite{devega08a,navarretebenlloch11a}, one atomic state can be tightly/loosely trapped playing the role of QEs (bath) and connect them through Raman lasers mimicking exactly the Hamiltonian $H_\mathrm{int}$ of Eq.~\ref{eqL:Hintt}, without any approximations (counter rotating processes do not appear) as we explain in more detail in Section~\ref{sec:experimental}. Interestingly, such cold atom implementation will allow one to obtain $g\sim O(J)$. Another platform where realistically tune the ratio $g/J$ almost arbitrarily is circuit QED, where superconducting qubits interact with coupled resonators~\cite{liu17a}. In this setup, one has to consider situations where the rotating wave approximation done in $H_\intt$ still holds.

For clarity, let us list altogether the assumptions/regime we are interested in along this manuscript:
\begin{itemize}
 \item We are interested in predictions in the continuum limit, i.e., $N\rightarrow \infty$, where finite size effects are negligible.
 \item We focus on the regime where the QE transition frequency lies within the band, i.e., $\Delta/J\in [-4,4]$.
 \item We illustrate the phenomenology for coupling constants $g\alt J$, for the sake of clarity of the figures. However, we typically provide analytical formulas for the regime $g/J\ll 1$ that allows one to understand the scaling of the features for weaker couplings. Notice that for the optical or circuit QED implementation we still demand that $g\sqrt{N} \ll \omega_a,\omega_e$ such that the counter-rotating terms of $H_\intt$ can be safely neglected and which also justifies considering a single band. For the purely atomic implementation~\cite{devega08a,navarretebenlloch11a}, this restriction can be relaxed.
 \item We are neglecting the coupling to other reservoirs because we want to capture the effects that arise merely from the interaction through the 2D structured bath. This coupling might introduce losses in the system, which timescale must be obviously larger than those of the phenomena we want to simulate.
 \item We study the problem of having one (or several) QEs prepared in a given initial state, $\ket{\Phi_0}_S$, while the bath is initially empty, $\ket{\mathrm{vac}}_B$, and then let the system evolve according to the dynamics dictated by the total Hamiltonian $H=H_S+H_B+H_\mathrm{int}$. Moreover, we restrict to cases where $\ket{\Phi_0}_S$ contains only a single excitation, as they already display the physics we want to explore.
\end{itemize}

\section{Theoretical framework \label{sec:theory}}

To analyze the QE dynamics we follow two alternative and complementary approaches: either integrating the Schr\"{o}dinger equation for the full QE-bath system numerically, or using analytical techniques~\cite{cohenbook92a} to solve exactly the QE dynamics.

\subsection{Numerical tools \label{sec:numerics}}

Even though we are interested in the limit $N\rightarrow \infty$, our numerical simulations always have a finite $N$. Thus, the main challenge to make a faithful numerical integration of the dynamics is to be able to simulate a large enough lattice such that finite size effects do not occur within the timescale of the simulation. There are three relevant scales in the system: i) the fastest timescale of the phenomena we want to simulate, e.g., spontaneous decay into the bath, that we denote in general as $\Gamma_{\mathrm{rel}}$, ii) the maximum speed of the propagation of excitations into the bath, that we denote as $v$, iii) the size of the system $N$. The rule of the thumb will be that the system size must satisfy $N> v \Gamma_{\mathrm{rel}}^{-1}$. 

The two-dimensional character makes it more challenging to simulate large system sizes as the number of bosonic modes scales with the area of the system ($N^2$). However, we exploit the simplified form of the different terms $H=H_S+H_B+H_\mathrm{int}$ to develop an efficient method that allows us to simulate our problem for system sizes $N>10^4$. As these methods can be extended to simulate other systems in higher dimensions and/or higher excitations we detail them here.

\emph{Spectral method}~\cite{press89a}.
This technique relies on the following fact: when written in $\kk$-space $H_B$ is diagonal, such that its generated time evolution is trivial to calculate, whereas $H_\mathrm{int}$ couples each QE to all $\kk$'s. In real space however, $H_\mathrm{int}$ has a simple form because each QEs couples to a single bosonic mode $a_{\nn_j}$, but the price to pay is that $H_B$ is non-diagonal. However, it is possible to combine the best of both representations to perform the simulation efficiently. The idea is to discretize our simulation time in steps $dt$, such that $t_n=n dt$, and perform the complete evolution of $H$ in four steps:
\begin{align}
 \ket{\Psi(t_{n+1})}=U_{\nn\rightarrow \kk}e^{-i H_\mathrm{int} dt}U_{\kk\rightarrow\nn}e^{-i (H_S+H_B) dt}\ket{\Psi(t_{n})}\,,
\end{align}
that is, we start with the wavefunction $\ket{\Psi(t_{i})}$ with the bath written in $\kk$-space, and apply the evolution of $H_S+H_B$ that can be precalculated analytically because the Hamiltonians are diagonal in that representation. Then, we make a change of basis in the bath modes to real space that we represent through $U_{\kk\rightarrow\nn}$ and apply the evolution of $H_\mathrm{int}$ which can also be precalculated analytically as it is a $2\times 2$ Hamiltonian for each QE-bath interaction. Finally, we change the basis again to $\kk$ space with $U_{\nn\rightarrow \kk}$ to prepare for the next step. The bottleneck of the simulation is the change of basis, as in principle it requires a number $O(N^4)$ of operations. However, by realizing that the change of basis is nothing but a Fourier Transform and by restricting to systems with $N=2^\alpha$ we can apply the Fast Fourier Transform algorithm which reduces the number of operations to $O(N^2\log(N))$. This method can be used for several QEs as well.

\emph{Discretizing frequency space.}
A further simplification can be obtained by realizing that the $\kk$ modes have a degeneracy in frequency space. The idea is to partition the frequency interval in small pieces of size $\delta\omega$, and just use the Hamiltonian with this discretized version which transforms it into a one-dimensional problem. Let us illustrate it with the single QE situation, but the results can be generalized for $N_e>1$. We discretize $[-4J,4J]$ in several frequencies $\omega_n$ with $\omega_n-\omega_{n-1}=\delta\omega$, such that we can rewrite $H_B$ and $H_\mathrm{int}$ as follows:
\begin{align}
 H_B&=\sum_\kk \omega(\kk) a_\kk^\dagger a_\kk=\sum_{\omega_n}\omega_n \Big(  a^\dagger_{\omega_n} a_{\omega_{n}}+ \sum_{\alpha=1}^{\tilde{N}(\omega_n)-1} a^\dagger_{\alpha,n}a_{\alpha,n}\Big)\,,\nonumber\\
 H_\mathrm{int}&=\frac{g}{N}\sum_\kk \left(a_{\kk} \sigma_{eg}+\mathrm{h.c.}\right)= \frac{g}{N}\sum_{\omega_n} \sqrt{\tilde{N}(\omega_n)} \left(a_{\omega_n}  \sigma_{eg}+\mathrm{h.c.}\right)
\end{align}
where $\tilde{N}(\omega_n)$ is the number of modes in the frequency interval $\omega_n$, proportional to the density of states of the bath. Notice, that the density of states can be calculated analytically in the thermodynamic limit for our models of interest (see below). We defined the operator $a_{\omega_n}$ that couples to the QE as follows:
\begin{equation}
 a_{\omega_n}=\frac{1}{\sqrt{\tilde{N}(\omega_n)}}\sum_{\kk=\kk(\omega_n)} a_\kk\,,
\end{equation}
and we denote the $\tilde{N}(\omega_n)-1$ orthogonal ones by the index $\alpha$ in $H_B$. As they do not couple to the QE they will not be populated and their dynamics can be ignored. This method can be extended to more QEs by simply taking into account the mode each QE is coupled to for each frequency.

\subsection{Analytical tools \label{sec:analytical}}

A typical approach to these problems consists of tracing the environment degrees of freedom and use the Born-Markov approximation to write an effective Master Equation \cite{gardiner_book00a} which describes the dynamics of the QEs. In this work, however, we use the resolvent operator technique~\cite{cohenbook92a}, that allows us to extrapolate between a perturbative description, which recovers the results of Born-Markov Master Equations, and a non-perturbative one when that description fails. 

Let us first illustrate it with the case of a single QE starting in an initial state, $\ket{\Phi_0}_S=\ket{e}$. As the total Hamiltonian $H$ conserves the number of excitations, $N_\mathrm{exc}=\sigma_{ee}+\sum_\kk a_\kk^\dagger a_\kk$, the combined wavefunction at any time can be written:
\begin{equation}
 \label{eqL:singleQEwave}
 \ket{\Psi(t)}=\left[C_e(t)\sigma_{eg}+\sum_\kk C_\kk a_\kk^\dagger \right]\ket{g}_S\otimes\ket{\mathrm{vac}}_B\,,
\end{equation}
where $C_{e}(0)=1$ and $C_\kk(0)=0$. Using resolvent operator techniques these probability amplitudes can be calculated as the (displaced) Fourier transform.
\begin{equation}
 \label{eqL:probabilityamp}
 C_\alpha(t)=-\frac{1}{2\pi i}\int_{-\infty}^\infty dE G_\alpha(E+i 0^+) e^{-i E t}
\end{equation}
for $t>0$ and where $G_\alpha(E+i 0^+)$ is the so-called retarded Green Function associated to the probability amplitude of the $\alpha$ state. By using $H$, it can be shown that in this case:
\begin{align}
 \label{eqL:Greenret}
 G_e(z)&=\frac{1}{z-\Delta-\Sigma_{e}(z)}\,,\\
 G_\kk(z)&=\frac{g}{\left(z-\omega(\kk)\right)\left(z-\Delta-\Sigma_{e}(z)\right)}\,,
\end{align}
where $\Sigma_e(z)=\frac{g^2}{N}\sum_\kk \frac{1}{z-\omega(\kk)}$ is the so-called self-energy, which in the continuum limit and above the real axis is given by:
\begin{align}
 \label{eqL:selfE}
 &\Sigma_e(E+i0^+)=\delta\omega_e(E)-i\frac{\Gamma_e(E)}{2}=\nonumber\\
 &=\frac{g^2}{(2\pi)^2}\iint d\kk \left[\mathrm{P.V} \frac{1}{E-\omega(\kk)}-i\pi\delta(E-\omega(\kk))\right]\,.
\end{align}

This function captures the effect of the bath on the dynamics of the QE. It can be separated in its real, $\delta\omega_e(E)$, and imaginary part, $\Gamma_e(E)$. The latter can be written in terms of the density of modes as follows $\Gamma_e(E)=2\pi g^2 D(E)$. A standard approximation in the literature consists of assuming that the coupling $g$ is sufficiently \emph{weak} such that $\Sigma_e(E+i0^+)\approx \Sigma_e(\Delta+i0^+)$. With that approximation the integral for $C_e(t)$ can be easily solved by applying Residue Theorem around the pole $z=\Delta+\Sigma_e(\Delta+i0^+)$ to obtain:
\begin{equation}
 \label{eqL:popsingMarkov}
 C_e(t)\approx e^{-i\left(\Delta+\delta\omega_e(\Delta)-i\frac{\Gamma_e(\Delta)}{2}\right)t}\,,
\end{equation}
which reproduces the expected behavior in the perturbative regime; that is, the coupling to the bath induces both a shift in the QE energy, $\delta\omega_e(\Delta)$, and an exponential decay of the population, with a decay rate given by $\Gamma_e(\Delta)$, that we denote from now on as $\delta\omega_M$ and $\Gamma_M$, respectively. This approach is commonly referred to as Wigner-Weisskopff or single-pole approximation, which predicts an exponential decay given by the Fermi's Golden Rule (FGR). 

Using that approximation, the integral for $C_\kk$ can also be solved applying Residue Theorem, obtaining:
\begin{align}
 \label{eqL:bathMark}
 C_\kk(t)&\approx g\Big[\frac{e^{-i\omega(\kk) t}}{\omega(\kk)-\Delta-\delta\omega_M+i\Gamma_M/2}\nonumber\\
 +&\frac{e^{-i\left(\Delta+\delta\omega_M-i\frac{\Gamma_M}{2}\right)t}}{\Delta-\omega(\kk)+\delta\omega_M-i\Gamma_M/2}\Big]\,,
\end{align}
where the first contribution comes with the pole at $E=\omega(\kk)$, whereas the last one comes from the one at $E=\Delta+\delta\omega_M-i\frac{\Gamma_M}{2}$ and which will ultimately vanish for $t\rightarrow \infty$ if $\Gamma_M\neq 0$. Eq.~\ref{eqL:bathMark} shows that, within the Markov approximation, the modes dominating the emission are the ones satisfying $\omega(\kk)\approx \Delta$.

However, the perturbative treatment may fail in a situation where $\Sigma_e(E)$ can not be considered a perturbation to $\Delta$ or when it suffers discontinuities and/or divergences as in our bath. In those cases, one must perform the exact Fourier integral of Eq.~\ref{eqL:probabilityamp} to obtain the correct dynamics as we will see in Sections~\ref{sec:1D} and \ref{sec:single}.

The situation with $N_e=2$ QEs coupled to the bath can be treated in an analogous way than the single QE situation when $\omega(\kk)=\omega(-\kk)$. In that case, the Hamiltonian $H_\intt$ can be written in terms of the $\sigma_{\pm}=\left(\sigma^1_{ge}\pm\sigma_{ge}^2\right)/\sqrt{2}$ as follows:
\begin{equation}
 H_{\mathrm{int}}=\frac{g\sqrt{2}}{N}\sum_{\kk^+,\alpha=\pm}\sqrt{1\pm\cos(\kk\cdot\nn_{12})} \left(a_{\kk,\alpha}^\dagger \sigma_{\alpha}+\mathrm{h.c.}\right)\,,
\end{equation}
where the $\kk$-sum has been restricted to the positive ones, i.e., $k_{x,y}>0$, and $a_{\kk,+}$ are the bath modes that couple to the symmetric and antisymmetric superposition given by:
\begin{align}
 \label{eq:akl}
 a_{\kk,\pm}=\frac{\left[(1\pm e^{i\kk\cdot\nn_{12}})a_\kk+ (1\pm e^{+i\kk\cdot\nn_{12}})a_{-\kk}\right]}{2\sqrt{1\pm \cos(\kk\cdot \nn_{12})}}\,,
\end{align}

It can be shown that these modes are orthogonal, i.e., $[a_{\kk,\alpha},a^\dagger_{\qq,\beta}]=\delta_{\alpha,\beta}\delta_{\kk,\qq}$, and that $H_{S,B}$ remain diagonal with those transformations. Thus, the dynamics of the symmetric/antisymmetric component is separable and can be calculated independently and in an analogue way to the single QE case. In particular, in Section~\ref{sec:two}, we focus on the dynamics starting in an initial state $\ket{\Phi_0}_S=\ket{\Phi_{\pm}}=\frac{1}{\sqrt{2}}\left(\sigma_{eg}^1\pm\sigma_{eg}^2\right)\ket{g,g}$, such that the probability amplitude to remain in $\ket{\Phi_{\pm}}$ is governed by a similar Green Function to that of a single QE Eq.~\ref{eqL:Greenret}, that we denote as $G_{\pm}(z)$, but replacing $\Sigma_e\rightarrow \Sigma_{\pm}=\Sigma_e\pm\Sigma_{12}$, where:
\begin{equation}
 \label{eqL:self12}
 \Sigma_{12}(z;\nn_{12})=\frac{g^2}{N^2}\sum_\kk \frac{e^{i\kk\cdot\nn_{12}}}{z-\omega(\kk)}\,.
\end{equation}

This function, $\Sigma_{12}(z;\nn_{12})$, accounts for the collective interaction mediated by the bath modes, which depends on the relative position $\nn_{12}$ between the two QEs. As it occurs with $\Sigma_e$, the $\Sigma_{12}(E+i0^+)=J_{12}(E)-i\frac{\Gamma_{12}(E)}{2}$ can also be separated in its real and imaginary components. The real one, $J_{12}$, is responsible for the coherent exchange of excitations between QEs, whereas the imaginary one renormalizes the decay rates of $\ket{\Phi_{\pm}}$ generating either super or subradiance depending on both the amplitude and phase of $\Gamma_{12}$.
 
 The situation with $N_e>2$ is in general not separable. However, in Section~\ref{sec:many} we will see a situation with $N_e=4$ where this separation is still possible. This means that for certain initial states the problem can be rephrased as the one of a single QE but with a modified self-energy.

\subsection{Effect of QE and bath losses \label{subsec:loss}}

Even though along the manuscript we focus on the QE dynamics induced only by the interaction with the bath, in practical situations both the QEs and the bath modes may couple to other reservoirs. These extra channels introduce losses at a rate $\Gamma^*$ and $\kappa$ for the QE and bath modes respectively. A way to account for such loss channels is to upgrade the Hamiltonian description to a density matrix one, in which the system-bath density matrix is governed by a Master Equation which reads~\cite{gardiner_book00a}:
\begin{align}
\label{eqL:master}
\frac{\partial \rho}{\partial t}=-i[H,\rho]+\frac{\kappa}{2}\sum_{\nn}\mathcal{L}_{a_\nn}[\rho]+\frac{\Gamma^*}{2}\sum_{j}\mathcal{L}_{\sigma^j_{ge}}[\rho]\,,
\end{align}
where we assume the loss rates $\kappa$ and $\Gamma^*$ to be the same for all the bath modes and QEs respectively and the Lindblad operator is defined as $\mathcal{L}_c[\rho]=(2c\rho c^\dagger -c^\dagger c\rho-\rho c^\dagger c)$. In the single excitation subspace, this equation can be formally solved as follows:
\begin{align}
\label{eqL:solution}
\rho(t)=\rho_0(t)+\rho_1(t)=e^{-i H_\mathrm{eff} t}\rho(0)e^{i H_\mathrm{eff}^\dagger t}+\int_0^t ds J\left[e^{-i H_\mathrm{eff} s}\rho(0)e^{i H_\mathrm{eff}^\dagger s}\right]\,,
\end{align}
where $\rho(0)=\ket{\Psi(0)}\bra{\Psi(0)}$ is the initial state. The term $\rho_0(t)$ is usually referred to as "no-jump" evolution and is governed by the effective non-Hermitian Hamiltonian, $H_\mathrm{eff}$, defined as $H_\mathrm{eff}=H-i\frac{\kappa}{2}\sum_\nn a^\dagger_\nn a_\nn-i \frac{\Gamma^*}{2}\sum_j \sigma^j_{ee}$. The other term, $\rho_1(t)$, describes the evolution where a quantum jump, described by $J[\cdot]=\sum_{\nn} a_\nn [\cdot]a^\dagger_\nn +\sum_{j} \sigma_{ge}^j  [\cdot]\sigma_{eg}^j$, has occurred. In the single excitation subspace, the state after a quantum jump occurs will always be $\ket{g\dots g}_S\otimes \ket{\mathrm{vac}}$ independent on whether the jump occurs at the QE or bath mode. Thus, the general solution can be expressed as:
\begin{align}
\label{eqL:solution2}
\rho(t)=\rho_0(t)+P_\mathrm{jump}(t) \ket{g\dots g}_S \bra{g\dots g}_S\otimes \ket{\mathrm{vac}} \bra{\mathrm{vac}}\,,
\end{align}

In the particular case where $\kappa=\Gamma^*\equiv \Gamma_\mathrm{loss}$, the effective non-hermitian Hamiltonian, $H_\mathrm{eff}=H-i\frac{\Gamma_\mathrm{loss}}{2}\hat{N}_\mathrm{exc}$, commutes with $H$. Thus, the solution can be easily written:
\begin{align}
\rho(t)&=e^{-\Gamma_\mathrm{loss} t} \ket{\Psi(t)}\bra{\Psi(t)}+\nonumber \\
&(1-e^{-\Gamma_\mathrm{loss} t})\ket{g\dots g}_S\bra{g\dots g}_S\otimes\ket{\mathrm{vac}}_B\bra{\mathrm{vac}}_B\,.
\end{align}
where $\ket{\Psi(t)}=e^{-i H t}\ket{\Psi(0)}$, i.e., the evolved state in the absence of losses. In Section~\ref{sec:experimental}, we will use these equations to estimate the impact of the losses in the observation of the phenomena emerging from the pure system-bath Hamiltonian $H$.

\section{1D structured reservoirs \label{sec:1D}}

In order to emphasize the new features obtained from the two-dimensional scenario compared to other type of reservoirs~\cite{bykov75a,john90a,kurizki90a,longui06a,longo10a,garmon13a,lombardo14a,redchenko14a,laakso14a,douglas15a,calajo16a,sanchezburillo16a,gonzaleztudela15c,shi16a} we revisit the one-dimensional model with nearest neighbour coupling used to describe waveguide QED setups. This model assumes hopping at a rate $J$ between $N$ bosonic modes distributed along a line. Thus, the bath Hamiltonian can be written in real and momentum space as follows:
\begin{equation}
 H_B=-J\sum_{\mean{n,m}} \left(a_n^\dagger a_m+\mathrm{h.c.}\right)=\sum_k \omega(k)a_k^\dagger a_k\,,
\end{equation}

As it occurs for the two-dimensional scenario, the bath Hamiltonian is diagonal in momentum space, giving rise to an energy dispersion $\omega(k)=-2J\cos(k)\in [-2J,2J]$. The dispersion is linear around the middle of the band, i.e., $\omega(\pm\frac{\pi}{2}+q)=\pm 2J\sin(q)\approx \pm 2J q$, whereas it is parabolic around the band edges; e.g., for the lower one, $\omega(k)\approx -2J\left(1-\frac{1}{2}k^2\right)$ for $k=|\kk|\ll 1$. This translates into a density of states which can be calculated analytically in this case 
\begin{equation}
 D(E)=\frac{1}{\pi\sqrt{4J^2-E^2}}\Theta(2J-|E|)\,,
\end{equation}
which shows a nearly constant density of states around the middle of the band $D(E)\approx \frac{1}{2\pi J}$, whereas it diverges as $1/\sqrt{|E|-2J}$ around the band edges.

\subsection{Single QE}

\begin{figure}
\centering
\includegraphics[width=0.78\linewidth]{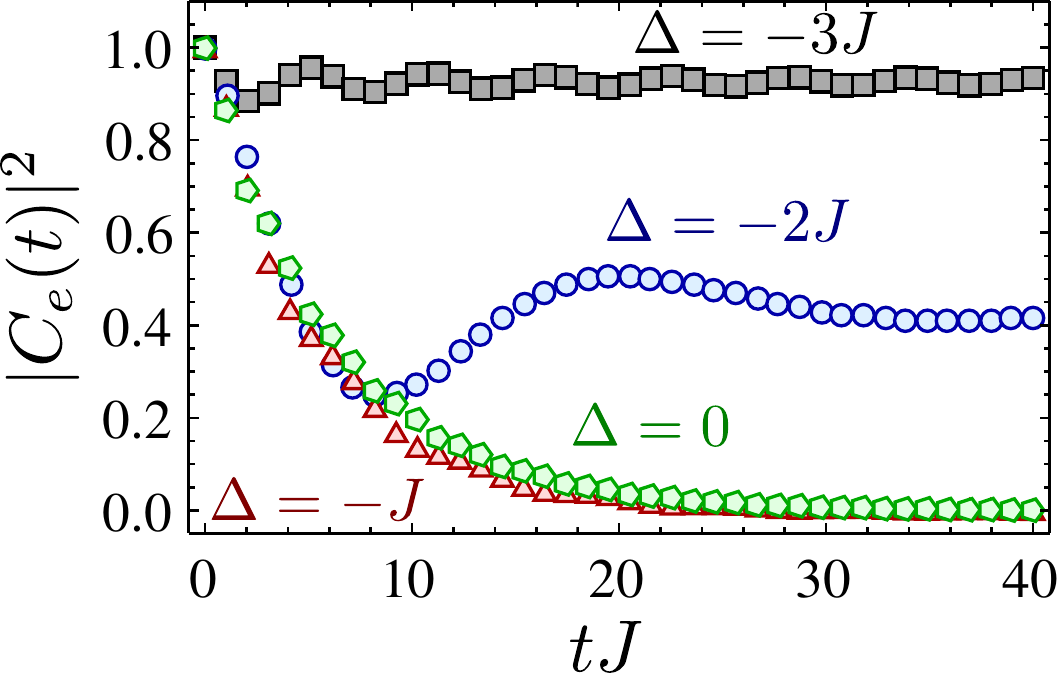}
\caption{Excited state population $|C_e(t)|^2$ for a single QE for $g/J=0.4$ interacting with a one-dimensional tight-binding bath for several $\Delta/J=-3,-2,-1,0$ as depicted in the legend.}
\label{fig1D1L}
\end{figure}

We consider the situation of a single QE that is initially excited, i.e., $\ket{\Phi_0}_S=\ket{e}$, and study its dynamics as a function of the detuning, $\Delta$, with respect to the central frequency. In Fig.~\ref{fig1D1L}, we plot the results from the exact integration of the evolution for several detunings with respect to the center of the band $\Delta/J=-3,-2,-1,0$ and $g=0.4J$. For detunings far from the band edge [$\Delta=-3J$], the spontaneous emission is quenched and the excitation remains mostly in the QE. For a detuning around the band edges, $\Delta=-2J$, the QE experiences the so-called fractional decay in which the excitations partly decay into the bath and partly remain localized. In contrast, in the center of the band, the QE experiences mostly an exponential decay of its population because of its interaction with the bath.

To gain analytical understanding we use the techniques described in Section \ref{sec:theory} which start out by calculating the self-energy for the one-dimensional bath. This can be done by taking the continuum limit and transforming the sum into an integral yielding:
\begin{align}
 \label{eqL:self1d}
 \Sigma_{e,\mathrm{1d}}(z)&=\frac{1}{2\pi}\int \frac{dk}{z-2J\cos(k)}=\nonumber\\
 &=\frac{1}{2\pi i}\oint \frac{dy}{-y^2+y z-1}=\pm \frac{g^2}{\sqrt{z^2-4 J^2}}\,.
\end{align}
where the $\pm$-sign depends on whether $\mathrm{Re}{z}\gtrless 0$. Within the Wigner Weisskopff approximation, the decay of the QE is basically given by $|C_e(t)|^2\approx e^{-\Gamma_e(\Delta)t}$, where: $\Gamma_e(\Delta)=2\pi g^2 D(\Delta)$. Thus, it predicts no decay for $\Delta\notin [-2J,2J]$ and a perfect exponential decay for $\Delta\in [-2J,2J]$. In Fig.~\ref{fig1D1L}, it is evident the limitation of the Markov or Wigner-Weisskopff approaches to capture the dynamics in the whole parameter space. In particular, close to the band edges, the QE does not decay completely, leading to the so-called fractional decay.

\begin{figure}
\centering
\includegraphics[width=0.78\linewidth]{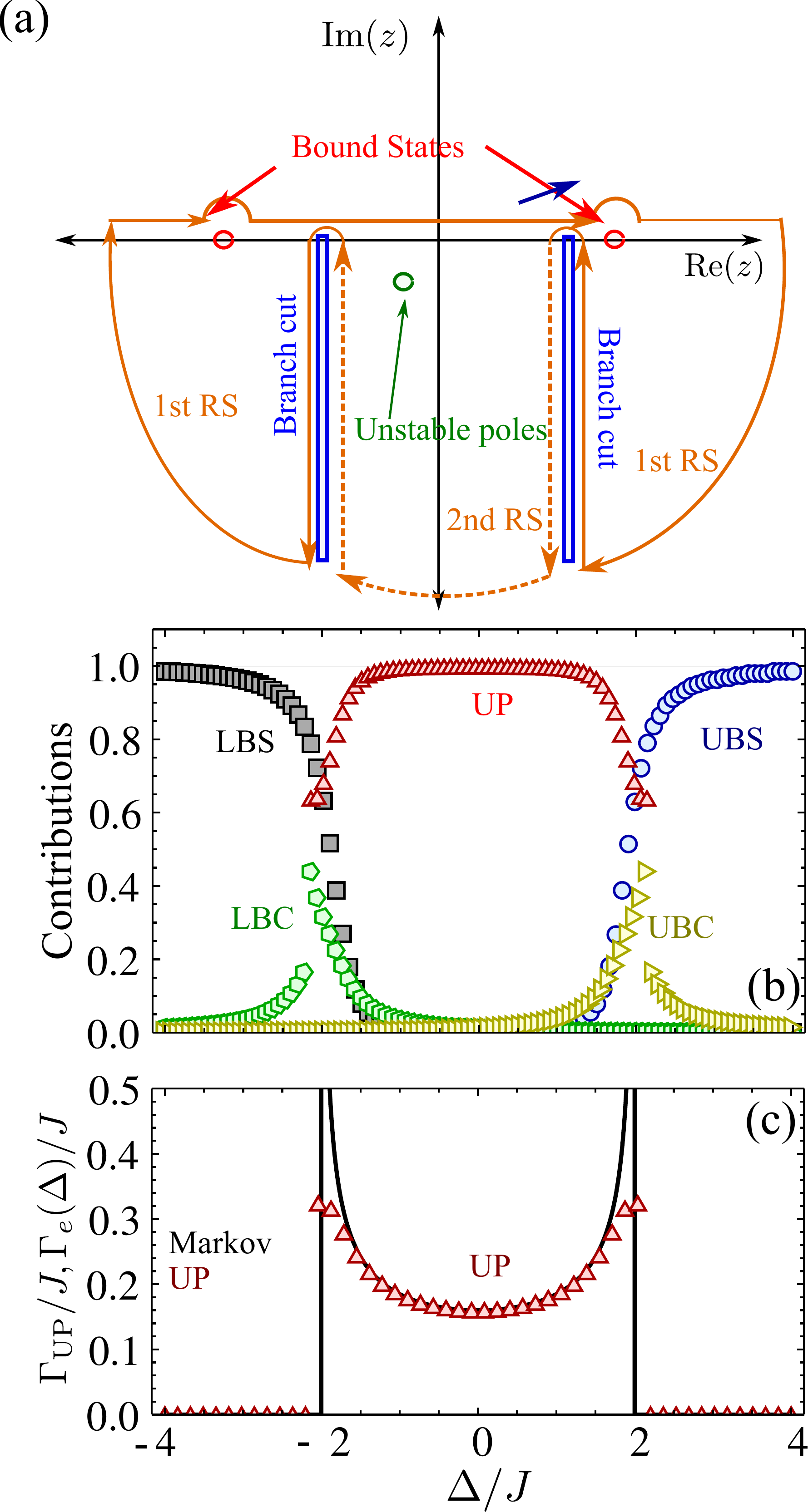}
\caption{(a) Contour of integration for 1D tight-binding bath taking two detours at the band edges $E=2J$ to avoid the branch cuts. (b) Contributions to the dynamics at $t=0$ for a situation with $g=0.4J$ as a function of $\Delta$. (b) Imaginary part of the complex pole (in the Second Riemann sheet) as a function of $\Delta$ compared to Markov prediction for $g=0.4J$.}
\label{fig1D2L}
\end{figure}

To go beyond the perturbative predictions it is convenient to calculate the Fourier transform appearing in $C_e(t)$ by closing the contour in the lower half plane and apply complex analysis techniques. The continuum $\omega(k)$ introduces branch cuts in $\Sigma_e(z)$, due to the presence of the $\sqrt{.}$ which tells us that we can not directly close the integral with a semicircle in the lower half plane. On the contrary, we decide to define the branch cuts of $\Sigma_{e,\mathrm{1d}}(z)$ at the band edges $z=\pm 2J-i x$, with $x>0$, such that to close continuously the contour we have to take a detour as depicted in Fig.~\ref{fig1D2L}(a).

\begin{figure*}
\centering
\includegraphics[width=0.78\linewidth]{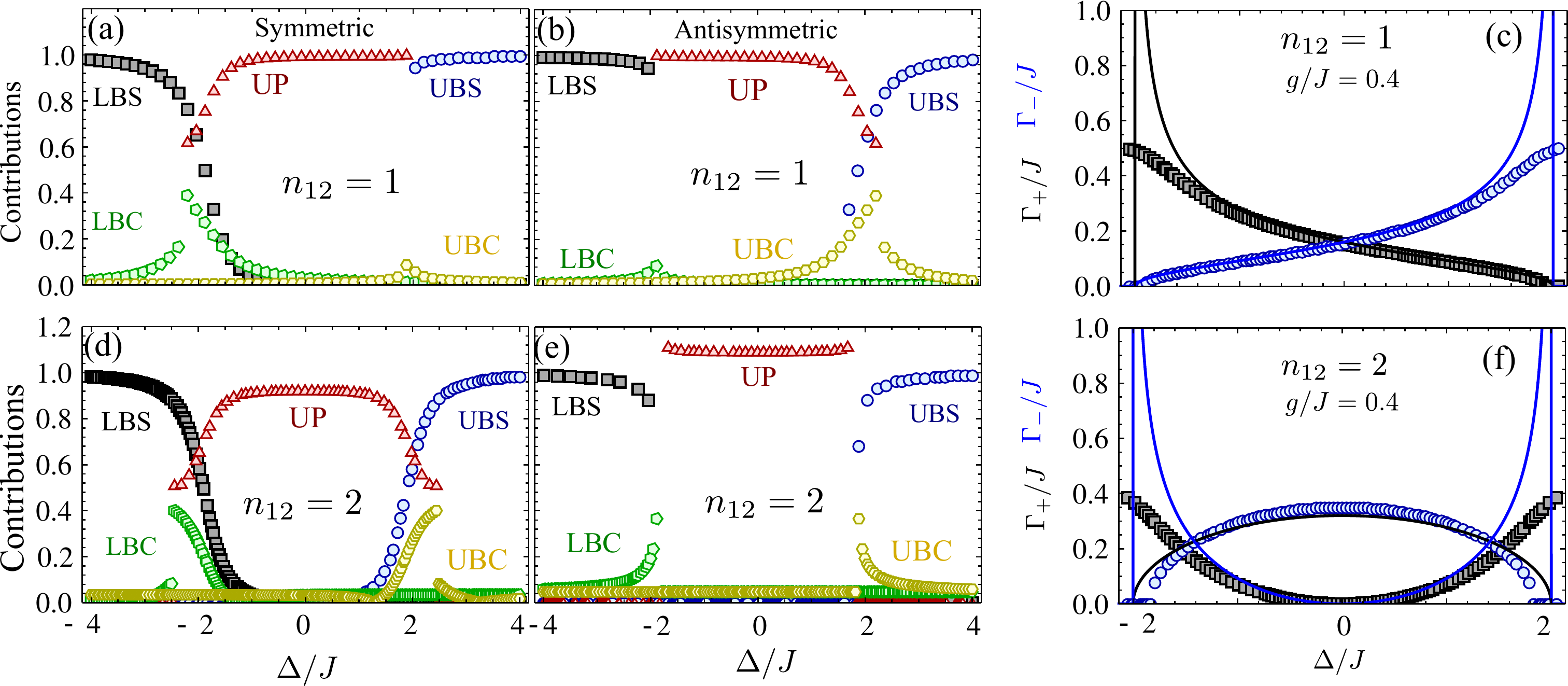}
\caption{(a-b) [(c-d)] Contributions to the dynamics at $t=0$ for the symmetric/antisymmetric component starting with a state $\ket{\Phi_{0,\pm}}$ for $g=0.4J$ and $n_{12}=1$ [$n_{12}=2$]. (c) [(f)] Comparison of the symmetric/antisymmetric decay rate as a function of $\Delta$ between the Markov prediction (solid) with the imaginary part obtained from the complex pole in the second Riemann sheet.}
\label{fig1D3L}
\end{figure*}

The dynamics can be calculated from the sum of different contributions, namely,
\begin{itemize}
 \item The contribution coming from real isolated poles of $G_e(E)$, which therefore satisfy:
 \begin{align}
  F(E)=E-\Delta-\Sigma_{e,\mathrm{1d}}(E)=0\,,
 \end{align}
 that emerge outside the continuum [in red in Fig.~\ref{fig1D2L}(a)]. These appear as a consequence of the so-called photon bound states (BS) and are responsible of the fractional decay observed in Fig.~\ref{fig1D1L}(a).
 \item The contribution appearing from the detour taken because of the branch cuts (BCs) at the lower/upper edges.
 \item When taking the detour at the band edges, the Green Function $G_e(z)$ has to be analytically continued to the second Riemann sheet, which can be easily done in this situation by replacing $\sqrt{.}\rightarrow -\sqrt{.}$ in $\Sigma_{e,\mathrm{1d}}$. This opens up the possibility of finding a complex pole of $F(E)$ which also contributes to the dynamics of $C_e(t)$. We denote this complex pole as \emph{unstable} pole (UP) because it is the one responsible for the spontaneous emission rate into the bath when $\Delta\in [-2J,2J]$
 \end{itemize}

 The rest of the elements introduced to close the contour can be shown to give no contribution to $C_e(t)$. Thus, the probability amplitude $C_e(t)$ can be finally written as a sum of the different terms
 \begin{align}
  C_e(t)=\sum_{\alpha=\mathrm{UBC,LBC}}C_\alpha(t)+\sum_{\beta=\mathrm{LBS,UBS,UP}} R_\beta e^{-i z_\beta t}\,,\nonumber
 \end{align}
where $R_\beta$ will be the residue of the real and unstable poles that we obtain through Residue Theorem and that gives the overlap of the initial wavefunction with the LBS/UBS/UP. For the single QE Green Function $\Sigma_e(z)$, it can be shown to be:
\begin{align}
\label{eqL:res}
 R_\beta=\left|\frac{1}{1-\partial_z\Sigma_{e,\mathrm{1d}}(z)}\right|_{z=z_\beta}\,,
\end{align}

In Fig.~\ref{fig1D2L}(b), we plot the contribution of the different elements to the dynamics at $t=0$ for $g=0.4J $ and scanning the detuning $\Delta$ in/out of the band. One can clearly distinguish the different regimes that explain the dynamics in Fig.~\ref{fig1D1L}; for QE energies well beyond the band, $|\Delta\pm2J|\gg g$, the LBS/UBS state contribution dominates with $|R_{\mathrm{LBS/UBS}}|\approx 1$. For regions deep in the band, $|\Delta|\ll J$, the dynamics are mostly governed by the UP contribution, which yields the exponential decay observed in Fig.~\ref{fig1D1L}. Finally, in the regions close to band edges, i.e., $|\Delta\pm 2J|\ll g$, the dynamics is a mixture between the BS, UP and the BC contribution. The latter leads to a power-law decay of the population scaling with $1/t^{3}$; however, the large overlap with the BS makes that the most visible feature is the fractional decay of the QE excitation and not the power law decay due to the BC.

Finally, in Fig.~\ref{fig1D2L} we compare the perturbative prediction of the decay rate, $\Gamma_e(\Delta)$ in solid black, with the one obtained from the imaginary part of the complex poles $2\mathrm{Im}z_{UP}$, showing how instead of diverging it arrives to a constant value which depends on the ratio $g/J$. 

Summing up, for $g\ll J$  and $\Delta$ lying within the band and far from band-edges, the dynamics follow the expected Markovian prediction of just an exponential decay with a rate given by FGR. Close to the band-edges $|\Delta\pm 2J|\ll g$, the Markovian prediction gets corrected because the QE does not decay completely because of the existence of a BS.

\subsection{Two QEs: super/subradiance}

The goal of this paper is to explore mainly the dissipative regime, that is, when the QE transition frequencies lie in the band, i.e., in 1D $\Delta\in [-2J,2J]$. In this regime, the most remarkable effect of the coupling of $N_e=2$ QEs to the bath is the renormalization of the lifetimes of certain atomic states. In particular, we will see that there are certain atomic states in which the decay is enhanced/suppressed because of the presence of other QEs coupled to the bath, that we label as super/subradiant states. We study these effects by considering an initial atomic state given by the symmetric/antisymmetric combination of a single excitation 
\begin{equation}
 \ket{\Phi_{0,\pm}}_S=\frac{1}{\sqrt{2}}\left(\sigma_{eg}^1\pm \sigma_{eg}^2\right)\ket{gg}\,.
\end{equation}

As we explain in Section~\ref{sec:theory}, the probability amplitude for the symmetric/antisymmetric component, $C_\pm(t)$, can be calculated in an analogous way to the single QE situation as they are separable and contain a single excitation. The collective symmetric/antisymmetric self-energy is given $\Sigma_{\pm,\mathrm{1d}}=\Sigma_{e,\mathrm{1d}}\pm \Sigma_{12,\mathrm{1d}}$, where $\Sigma_{12,\mathrm{1d}}$ (in the first Riemann sheet) reads
\begin{align}
 \Sigma_{12,\mathrm{1d}}(z;n_{12})&=\frac{1}{2\pi}\int_{-\pi}^\pi dk \frac{e^{i k n_{12}}}{z+2J\cos(k)}\nonumber\\
 &= \pm i\frac{g^2}{\sqrt{4J^2-z^2}}\left(-\frac{z}{2J}\mp i\sqrt{1-\left(\frac{z}{2J}\right)^2}\right)^{n_{12}}
\end{align}
where the $\pm$ sign corresponds to the situation where $\mathrm{Re}(z)\lessgtr 0$ and $n_{12}$ is the relative position between the QEs in the lattice. As we have the complete analytical behaviour of $\Sigma_{\pm,\mathrm{1d}}(z)$, we make a similar study of the different contributions to the dynamics of $C_{\pm}(0)$ as we did for the single QE following a similar contour of integration as described in Fig.~\ref{fig1D2L}(a). The summary of the results are shown in Fig.~\ref{fig1D3L} for two different distances, $n_{12}=1,2$, and for the symmetric and antisymmetric components. We also compare in panels (c,f) the exact imaginary component of the complex pole compared to the Markov prediction. Let us summarize the main particularities of the two QE situation, with respect to the single QE one:
\begin{itemize}
 \item In the lower edge, the LBS contribution survives for the whole parameter range for both $n_{12}$'s. This can be shown to be the case for all $n_{12}$ as $\Sigma_{+,\mathrm{1d}}(z=-2J;n_{12})\rightarrow \infty $, which is a necessary condition for finding such bound states \cite{shi16a}. Interestingly, the contribution of the antisymmetric LBS vanishes at a critical $\Delta$ because the divergences from $\Sigma_{e,\mathrm{1d}}$ and $\Sigma_{12,\mathrm{1d}}$ compensate, leading to a critical detuning from the upper edge $\Delta_c=\lim_{E\rightarrow -2J-0^+}\Sigma_{-,\mathrm{1d}}(E;n_{12})=\frac{g^2 n_{12}}{2 J}$ to observe the antisymmetric LBS contribution.
 
 \item In the upper edge, however, the merging of the BS into the continuum alternates between the symmetric and antisymmetric component; that is, for odd [even] $n_{12}$ the symmetric [antisymmetric] UBS contribution disappears at an analog critical detuning from the upper edge~\cite{calajo16a}.
 
 \item In the middle of the band, the UP contribution (in red) dominates in both the symmetric and antisymmetric components. As shown in panels (c,f), depending on the relative value between $\Delta$ and $n_{12}$, the decay rates of the symmetric/antisymmetric components get enhanced/suppressed with respect to the single QE situation. This is what we will label in this paper as super/subradiance to settle the nomenclature. Moreover, we show that there are certain values in which the imaginary part can be perfectly suppressed, e.g., $\Delta=0$ and $n_{12}=2$, in which the unstable pole becomes stable, that we denote as perfect subradiance; There exists also the opposite situation where $\Gamma_{\pm}/\Gamma_e=2$; we will refer to the corresponding $\ket{\Psi_{\pm}}$ as superradiant states.
 
\end{itemize}

As this manuscript is focused on the dynamics emerging when the QEs are spectrally tuned within the band, let us explore in more detail the phenomenon of super/subradiance as it will be important for the discussion of the two-dimensional scenario. Let us initially apply the Markovian approximation to get an intuition on which values super/subradiant might appear:
\begin{equation}
 \label{eqL:1d}
 \Gamma_{\pm}(\Delta)\approx \Gamma_e(\Delta)\left[1\pm \cos(k(\Delta) n_{12})\right]\,
\end{equation}
where $k(\Delta)=\arccos\left(-\frac{\Delta}{2J}\right)$ is the wavelength associated to the propagating bath modes at the frequency of the QEs. Therefore, for distances satisfying $\cos\left(k(\Delta)n_{12}\right)=1$ the symmetric [antisymmetric] will be perfectly super[sub]radiant and viceversa for $\cos\left(k(\Delta)n_{12}\right)=-1$, which is what we observe in Fig.~\ref{fig1D3L}. 

\begin{figure}[!t]
\centering
\includegraphics[width=0.78\linewidth]{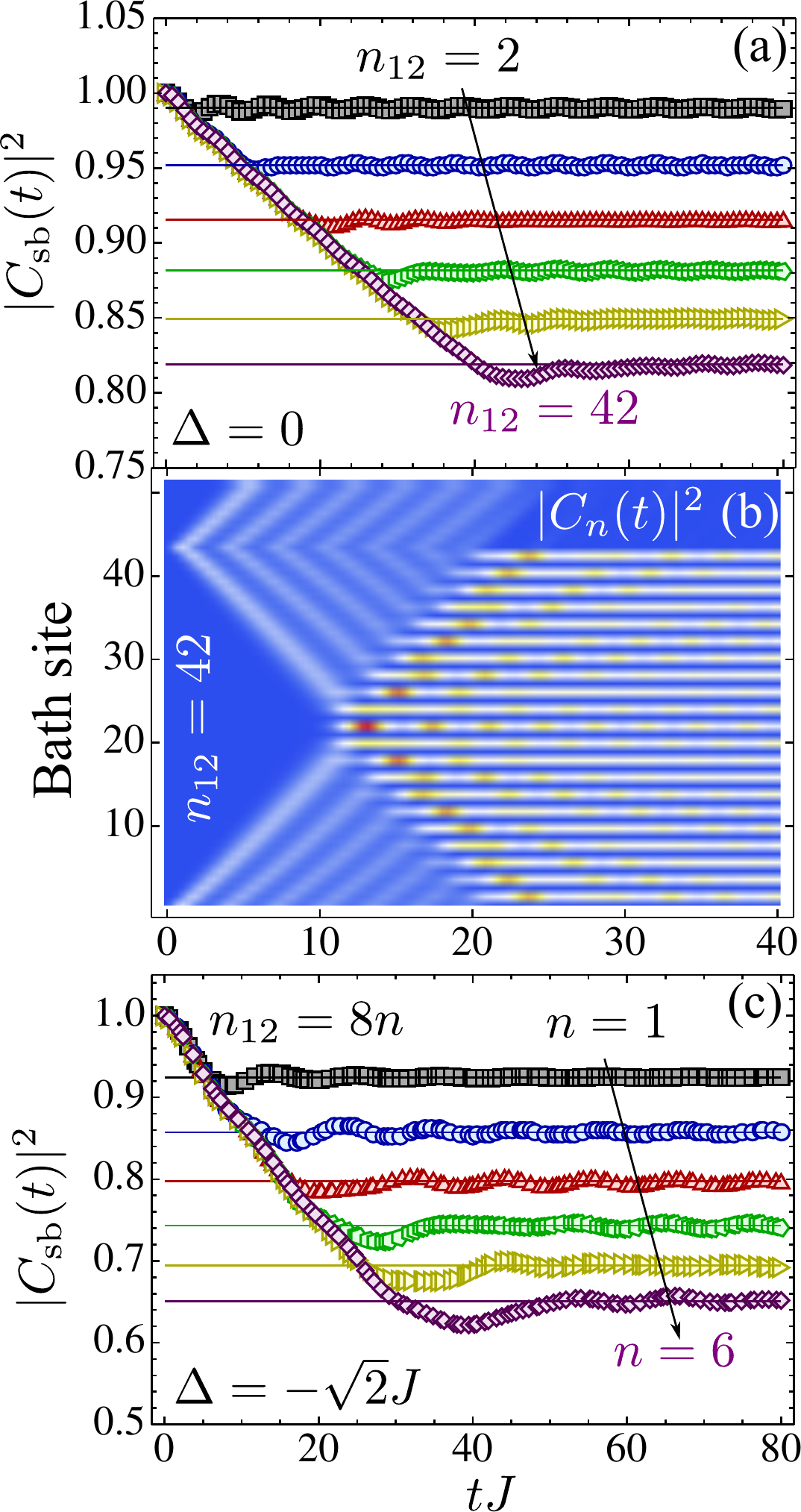}
\caption{(a) [(c)] Dynamics for the population of subradiant states $|C_\mathrm{sb}(t)|^2$ for $g=0.1J$ and $\Delta/J=0 [-\sqrt{2}]$ and several distances as depicted in the legend. (b) Bath population dynamics $|C_n(t)|^2$ as a function of time for a situation with $\Delta=0$, $g=0.1J$ and $n_{12}=42$.} 
\label{fig1D4L}
\end{figure}

A legitimate question is whether these perfect super/subradiant states survive beyond the Markovian approximation and what are the corrections to it. Let us illustrate the result for the subradiant configuration and, e.g., focusing on what happens to the asymmetric state contribution $G_{-}(z)$.  The exact pole equation we need to solve in this case reads:
\begin{equation}
 z-\Delta=\frac{g^2}{\pi}\int_0^\pi dk \frac{1-\cos(k n_{12})}{z+2J\cos(k)}\,.
\end{equation}

In general, the integral appearing in the right-hand side of the equation diverges. However, it can be shown that by choosing the QE positions such that $\cos(k(\Delta) n_{12})=1$, and $z=\Delta$ the integral vanishes because of symmetry arguments and because the divergence is cured as can be shown by expanding the integrand around $k(\Delta)+q$ for $q\ll 1$:
\begin{equation}
 \frac{1-\cos\left(\frac{2 k \pi}{k(\Delta))}\right)}{\Delta+2J\cos(k)}\Big|_{k=k(\Delta)+q}\approx\frac{q}{2\sqrt{1-\frac{\Delta^2}{4J^2}}}\rightarrow 0\,,
\end{equation}

This proves that $z_\mathrm{sb}=\Delta\in \mathbb{R}$ is indeed a solution of the pole equation for such distances even beyond the  Markovian approximation. The only thing left to prove is that the associated residue, $R_{\mathrm{sb}}$, of such solution is not zero. This residue is directly related to the excitation that remains within the QEs in the long time limit, i.e., $C_\mathrm{sb}(\infty)= R_{\mathrm{sb}}$, and is given by:
\begin{align}
  R_{\mathrm{sb}}=\frac{1}{1-\partial_z\Sigma_{-}(z)|_{z=z_\mathrm{sb}}}=\frac{1}{1+n_{12}\frac{\Gamma_e(\Delta)}{2 |v_g(\Delta)|}}.\,,
\end{align}
where $v_g(\Delta)=\partial_k\omega(k)|_{k=k(\Delta)}=\sqrt{4J^2-\Delta^2}$ is the group velocity at frequency $\Delta$.
Thus, the overlap with the initial state is mostly $1$, except from a retardation effect which takes into account the time, $n_{12}/(2v_g(\Delta))$, it takes to the bath excitations to move among the QEs compared to the decay time into the bath $\Gamma_e(\Delta)$. We numerically certified these predictions in Fig.~\ref{fig1D4L}, where we plot the subradiant state dynamics for a situation with $g=0.1J$ and $\Delta=0$ (in the middle of the band) and $\Delta=-\sqrt{2}J$. In both cases, we observe how the subradiant state decays in a timescale $\Gamma_e(\Delta)$ until the excitation from one QE arrives to the other. At that moment a destructive interference between the bath excitations from both QEs occurs quenching the decay into the environment. The longer the separation and the smaller the group velocity, the larger is the retardation effect. However, we emphasize that the imaginary part of the pole is still strictly zero, and it is only the overlap with the initial state the one that gets corrected. Moreover,
 in Fig.~\ref{fig1D4L}(b) we plot the bath population dynamics for a situation with $\Delta=0$ and $n_{12}=42$ where it is clearly illustrated the interference effect, which not only quenches the spontaneous emission into the environment, but also localizes the bath excitations between the QEs.

The collective phenomena discussed here have a very intuitive explanation in the context we are studying here and for $g\ll J$ and $\Delta=0$. In that case, energy conservation $\omega(\kk)\approx \Delta=0$ imposes that in the long time limit, the only $k$-modes that can be populated have momenta around $\pm k(\Delta)$. If we restrict the Hamiltonian $H_\mathrm{int}$ to those modes, we see that the subradiant state decouples continuously, whereas the superradiant coupling gets a factor of $\sqrt{2}$. Furthermore, the photons emitted in the reservoir at those $k$ coming from the two QEs interfere, either destructively or constructively depending on the phase (sign) of the superposition in the initial state, giving rise to the sub and superradiance. This interference exists as long as there is no which-way information indicating which QE has emitted the bath mode. Since the temporal size of the wavepacket emitted by each of them is of the order of $t_g=1/\Gamma_e(\Delta)$ and it propagates at velocity $v_g(\
Delta)$, there will be 
which-way information whenever the distance between the QEs is of the order $v_g t_g$ that will make the interference disappear.

\begin{figure*}
\centering
\includegraphics[width=0.9\linewidth]{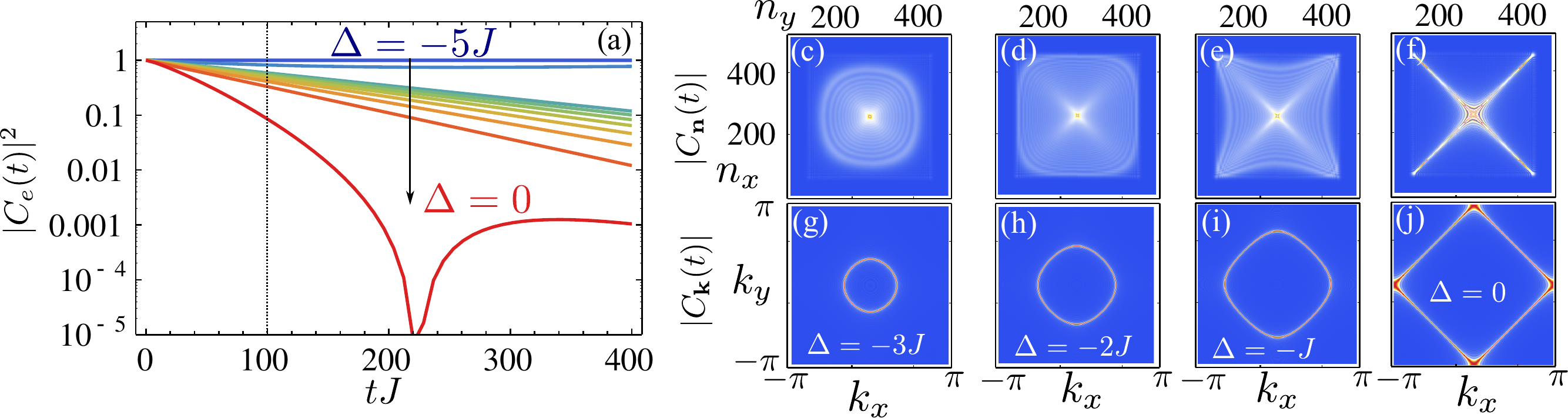}
\caption{(a) Excited state population $|C_e(t)|^2$ for a single QE for $g/J=0.1$ for $11$ equally spaced $\Delta$'s from $-5J$ to $0$. (c-f) [(g-j)] Absolute value of the probability amplitude of the bath modes for positions $\nn=(n_x,n_y)$ [momenta $\kk$] at a time $tJ =100$ for $\Delta/J=-3,-2,-1,0$.}
\label{fig2L}
\end{figure*}

\section{Single QE dynamic for two-dimensional baths \label{sec:single}}

After having analyzed the dynamics of one-dimensional structured reservoirs, let us explore what phenomena emerge from the two-dimensional bath that we described in Section~\ref{sec:system}.
As we do in the main manuscript \cite{gonzaleztudela17b}, we first proceed to solve the evolution of an initially excited QE numerically to see what features can be observed. Then, we try to understand them through the Wigner-Weisskopff approach. Finally, we perform the exact integration of Eq.~\ref{eqL:probabilityamp} to understand the different contributions which give rise to the dynamics.

In Fig.~\ref{fig2L}(a), we show the evolution of an initially excited QE coupled with $g/J=0.1$ for different $\Delta$'s ranging from $\Delta=-5J$, where the QE energy lies far from the band, to $\Delta=0$, which corresponds exactly to the energy at the middle of the band. Moreover, we complement this figure with panels (b-i) where we plot the bath population in position (upper row) and momentum space (lower row) at time $tJ=100$ for $\Delta/J=-3,-2,-1,0$. The QE goes from a situation where it is not emitting, i.e., $\Delta=-5J$, to situations where the QE emits into the bath modes at a faster timescale as we move closer to the middle of the band. At the band edge, i.e., $\Delta\approx -4 J$, the QE shows what has been called as fractional decay as it occurs for one-dimensional systems. Along this manuscript we are more interested in what happens in the middle of the band, which has no analogue with other types of reservoirs, where we observe: i) at short times the QE decays exponentially, but with a 
timescale which is 
not the one obtained from perturbative treatments as it predicts an infinitely fast decay; ii) at longer times the QE shows an oscillation followed by a subexponential relaxation; iii) as shown in Figs.~\ref{fig2L}(f,j), the emission into the bath becomes highly anisotropic, ultimately decaying as if there were just two quasi-1D modes.

\subsection{Wigner-Weisskopff approach}

As we show in Section \ref{sec:theory}, in both the exact and the Wigner Weisskopff approach, the function which dominates the dynamics of a single QE coupled to a bath is the so-called self-energy $\Sigma_e(z)$ defined in Eq.~\ref{eqL:selfE}. Interestingly, for the simplified 2D bath that we are considering, this self-energy can be calculated analytically, obtaining
\begin{align}
\label{eqL:self2D}
  \Sigma_{e}(z)=\frac{2 g^2}{\pi z}K\left[\left(\frac{4J}{z}\right)^2\right]\,,
\end{align}
where $K(m)$ is the complete elliptical integral of the first kind, using the following convention:
\begin{align}
 K(m)=\int_0^{\pi/2} d\phi \frac{1}{\sqrt{1-m \sin^2(\phi)}}\,.
\end{align}

This function is purely real when $m\in(-\infty,1)$, which means that $\Sigma_{e}(z) \in \mathbb{R}$ if $\mathrm{Re}(z)\notin(-4J,4J)$, whereas it is complex if $\mathrm{Re}(z)\in [-4J,4J]$. Within the Wigner-Weisskopff approach, the dynamics of the probability amplitude is just dominated by the values of the self energy close to the real axis, $\Sigma_e(\Delta+i 0^+)$, which real and imaginary part we plot in Fig.~\ref{fig1L}(a). In particular, if we are only interested in the population decay, Wigner Weisskopff approach just predicts an exponential decay of the population $|C_e(t)|^2\approx e^{-\Gamma_M t }$ given by the decay rate predicted by FGR, $\Gamma_M=\Gamma_e(\Delta)$. Therefore, this approach fails to approximate both the regions close to the band edge, i.e., $\Delta/J\approx \pm 4$ and the reversible dynamics and slow relaxation in the middle of the band, $\Delta/J\approx 0$.

Regarding the population of the bath modes, the Markov approximation given by Eq.~\ref{eqL:bathMark} rightly predicts that within the band, where $\Gamma_e(\Delta)$ is finite, the population is dominated by the $\kk$ modes which satisfy $\Delta\approx \omega(\kk)$. However, it fails in the middle of the band, as the $\Gamma_M\rightarrow \infty$ predicts that the populations of the modes will be $0$. Nevertheless, when looking at panels (f), (j) we observe that the modes along the lines $\pm k_x\pm k_y=\pm \pi$ are indeed populated. This directional emission into the bath can be considered as the 2D analogue of chiral emission 1D waveguides~\cite{lodahl16a,mitsch14a,sollner15a}, as the expected $2\pi$-angle emission transforms into a more restricted emission distribution. Therefore, it deserves special attention to characterize it properly and explore what new effects emerge from it.

\subsection{Exact integration}

\begin{figure}
\centering
\includegraphics[width=0.8\linewidth]{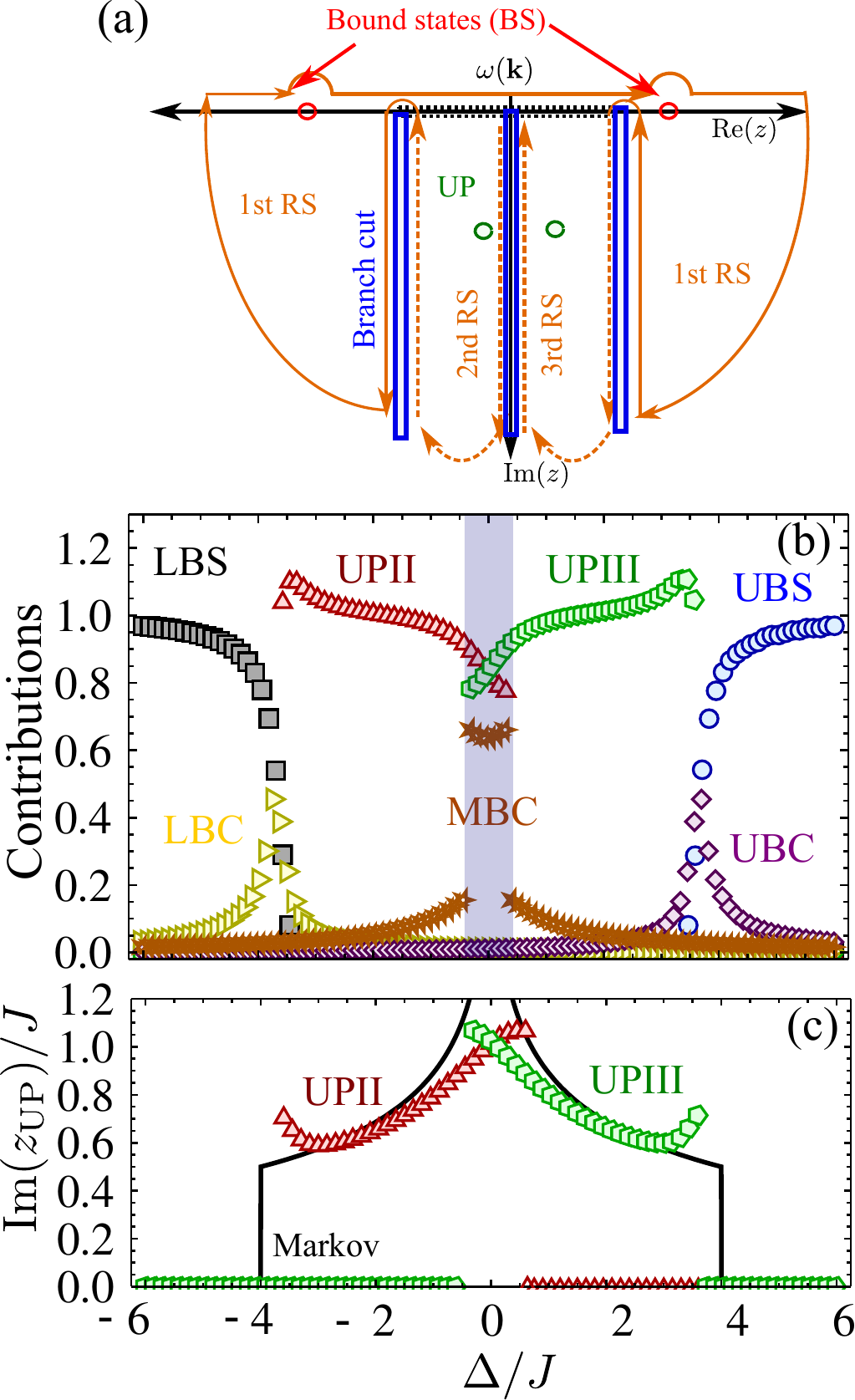}
\caption{(a) Contour of integration to calculate $C_e(t)$ from Eq.~\ref{eqL:probabilityamp}. When closing the contour from below, it crosses three branch cuts $G_e(z)$, the one of the band edges, and the extra branch cut that appear in the middle of the band. (b) Different contributions to the dynamics of the QE population at $t=0$ for $g=J$ as depicted in the legend. (c) Decay rate, $\Gamma_e(\Delta)$, as a function of $\Delta/J$ associated to the UPs of the analytical continuation of the Green Function $G_e(z)$. The markers correspond to the numerical exact calculation while the solid ones are the Markov predictions.}
\label{fig4L}
\end{figure}

The integration of the probability amplitude $C_e(t)$ can be done exactly by using complex analysis integral techniques as we already explained for the one-dimensional bath. In particular, we would like to close the integration path to find the poles of $G_e(z)$ and apply the Residue Theorem to calculate their contributions. As $t>0$, we have to close the contour through the lower part of the complex plane, i.e., $\mathrm{Im}{z}\rightarrow -\infty$. However, when closing the contour we face several difficulties as explained schematically in Fig.~\ref{fig4L}(a), namely,:
\begin{itemize}
 \item We have already pointed out that the elliptical integral appearing in the $\Sigma_e(z)$ displays a discontinuity at the band edges, i.e., $\mathrm{Re}{z}/J=\pm 4$. This is the typical discontinuity that appears in standard quantum optical problems because of band edges \cite{cohenbook92a} or also in a 1D waveguide as we explained in Section~\ref{sec:1D}. This discontinuity is associated to a BC of $\Sigma_e(z)$ (and therefore of $G_e(z)$) that we can not cross with our integration path. The solution consists of taking a detour of our integration path as depicted in Fig.~\ref{fig4L}(a), and use the analytical continuation in the second/third Riemann sheets. 
 
 \item The 2D bath, however, introduces a new variant which, up to our knowledge, has not been considered before in the quantum optical scenario. It can be shown that the elliptic integral also has another BC \emph{in the middle of the band}. This forces us to take an extra detour in our integration contour as depicted in Fig.~\ref{fig4L}(a), appearing at the third Riemann sheet. The analytical continuation of $\Sigma_e(z)$ in the second/third Riemann sheet can be obtained by using the properties of the elliptic integrals~\cite{abramowitz66a,morita71a}:
 \begin{align}
  \label{eqL:selfIIandIII}
  \Sigma_{e}^{\mathrm{II [III]}}(z)=\frac{2g^2}{\pi z}\left[K\left[\left(\frac{4J}{z}\right)^2\right]\pm2i K\left[1-\left(\frac{4J}{z}\right)^2\right]\right]
 \end{align}
 which should be used when $\mathrm{Re}(z)\in (-4J,0) [(0,4J)]$ respectively. The choice of this linear combination for the elliptic integrals is done to guarantee the continuity of the self-energy along the contour of integration.
 \end{itemize}

Therefore, the dynamics of $C_e(t)$ can be separated in different contributions:
\begin{itemize}
 \item First, we need to find the poles in the real axis that emerge out of the continuum that correspond to BS's. As the self-energy diverges at the band edges, i.e., $\Sigma_e(\pm 4J)\rightarrow \infty$, the BSs exist in all parameter regime, $(g/J,\Delta/J)$, as was shown in Ref.~\cite{calajo16a,shi16a}. These poles (and their corresponding residues) are obtained by solving the pole equation (and applying Residue Theorem respectively):
 \begin{align}
  \label{eqL:BSe}
  E_{\BS}&=\Delta-\Sigma_e(E_\BS)\,,\\
  R_{\BS}&=\frac{1}{1-\partial_E\Sigma_e(E)}\Big|_{E=E_\BS}\,.
 \end{align}

 As we have two band edges, we always find an upper and a lower BS that we denote as UBS/LBS. 
  
 \item Secondly, we also need to find solutions of the pole Equation in the second/ third Riemann sheet of the analytical continuation of $G_e(z)$, which can be done by solving Eq.~\ref{eqL:BSe}, but replacing $\Sigma_{e}\rightarrow \Sigma_{e}^{\mathrm{II,III}}$ respectively. These poles are generally complex, and therefore lead to the exponential decay of population, which is why they are usually called \emph{unstable} poles. We denote by UPII/UPIII to the unstable poles in the second/third Riemann sheet respectively.
 
 \item Finally, we need to add up the contributions coming from the detour introduced by the BC of $\Sigma_e(z)$. In this case we have three of them that we denote by LBC/MBC/UBC for the lower/middle/upper BC contributions. 
 
 \item The rest of the contributions coming from the circular detours around the edges and the bigger one to close the contour can be shown to be zero.
\end{itemize}

In Fig.~\ref{fig4L}(b) we plot the absolute value of the different aforementioned contributions to the dynamics of $C_e(t)$ at $t=0$ of a single QE for a coupling $g=J$ and as a function of $\Delta/J$. The absolute value of some of the contributions, e.g., UPs, can be larger than $1$ because they acquire a non negligible imaginary part. However, when summing them up they all interfere arriving to $C_e(0)=1$. Let us highlight the main findings: 

\emph{i)} First, we show that the UBS\&LBS survive for all $\Delta$ regimes, including values deep inside the band. In fact, it is possible to find a good approximation to the energies of the UBS/LBS for a wide range of $g$'s, which reads:
\begin{align}
E_{\mathrm{LBS}}&=-4J-\frac{g^2}{4\pi J} W\left(\frac{128 J^2\pi}{g^2} e^{-4 J\pi (\Delta+4J)/g^2}\right)\,,\\
E_{\mathrm{UBS}}&=4J+\frac{g^2}{4\pi J} W\left(\frac{128 J^2\pi}{g^2} e^{-4 J\pi(4J- \Delta)/g^2}\right)\,,
\end{align}
where $W(x)$ is the so-called product-log or Lambert function~\cite{abramowitz66a}. Notice  that in the strongly non-perturbative regime: $|\Delta\pm 4J|\ll g^2/(4\pi J)$, for the lower bound states the energy is given: $E_{\mathrm{LBS}}-4J\approx -\frac{g^2}{4\pi J} W\left(\frac{128 J^2\pi}{g^2} \right)\approx -\frac{g^2}{4\pi J}\log\Big(\frac{128 J^2\pi}{g^2}\Big) $.

\emph{ii)} Around the band-edges $|\Delta|\approx 4J$, the main contribution to the dynamics is given by both BS and BC. As it is well known, the BC contribution leads to subexponential decays. However they are typically hidden by the BS contribution which gives rise to the \emph{fractional decay} of Fig.~\ref{fig2L}(a) \cite{john90a,kurizki90a}.

\begin{figure}
\centering
\includegraphics[width=0.8\linewidth]{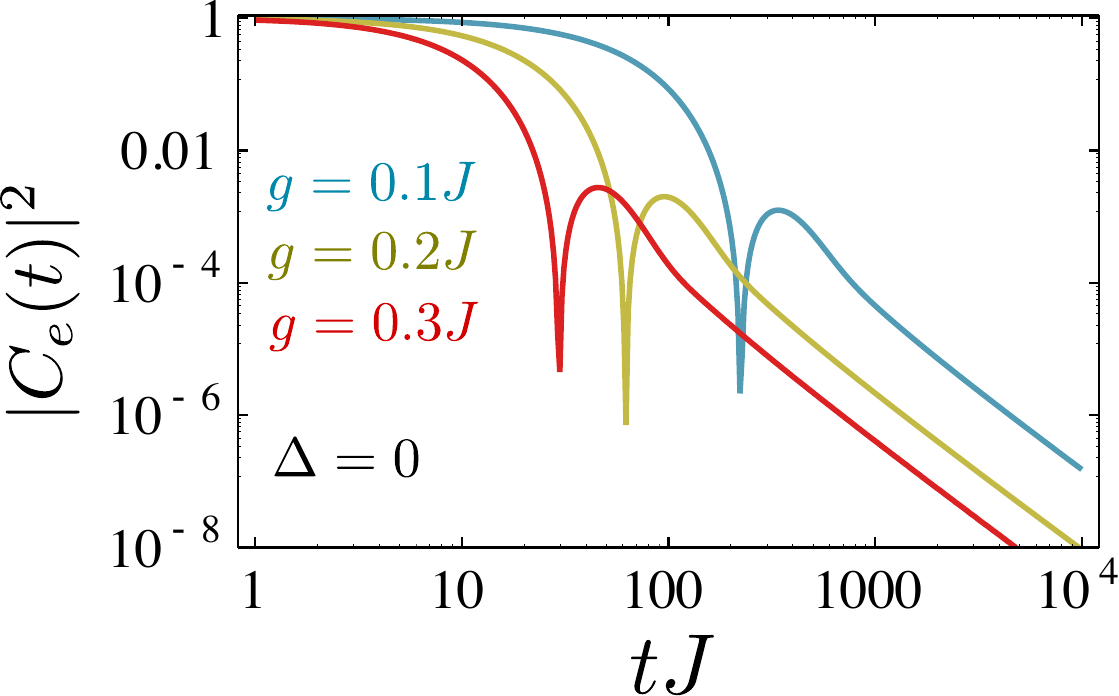}
\caption{Excited state population $|C_e(t)|^2$ in logarithmic scale for $\Delta=0$ and different $g/J=0.1,0.2,0.3$ as depicted in the legend.}
\label{fig5L}
\end{figure}

\emph{iii)} The most interesting region for our model lies within the band. In particular, we see that because of the existence of the MBC there is a range of energies, $\Delta\in [-\frac{g^2}{2J},\frac{g^2}{2J}]$, where two UP's from the second/third Riemann sheet can be found. The coexistence of both solutions can be traced back to the finite value of $\delta\omega_e(E)$ around $E\lessgtr 0$, which \emph{pushes} the energy of the UP to remain in the second/third Riemann sheet even for $\Delta\gtrless 0$. In Fig.~\ref{fig4L}(c), we plot both the imaginary part of such UP's showing that they have similar imaginary part (the same at $\Delta=0$, whereas different real energy (not shown) which difference provides the period of the oscillation shown in Figs.~\ref{fig2L}(a) and \ref{fig5L}. However, as the imaginary part and the real one scale in the same way with $g$ and are of the same order, the oscillations will be \emph{overdamped}, and this is why we only observe a single oscillation. From this analysis,  
we can even find an approximated expression of the imaginary part of the UPS at $\Delta=0$, which tells us how the Markovian prediction $\Gamma_M(0)\rightarrow \infty$ renormalizes to Non-Markovian decay rate which reads:
\begin{equation}
\label{eqL:singGam}
 \bar{\Gamma}_e\approx \frac{g^2}{\pi J}W\left(\frac{32\pi J^2}{g^2}\right)\approx \frac{g^2}{\pi J}\log\left(\frac{32\pi J^2}{g^2}\right)\,.
\end{equation}
where in the last equality we assumed $g\ll J$. This is an interesting result as it shows: i) that exponential relaxation into the bath can occur in places where perturbative approaches fail; and ii) it occurs at a rate $O\left(\frac{g^2}{J}\log(\frac{J^2}{g^2})\right)$ rather than the standard $O\left(g^2/J\right)$ obtained from FGR. Importantly, the coexistence of both UP's is also accompanied by a sudden jump in the MBC contribution. Thus, when the UP's have decayed we can also observe the slow relaxation dynamics given the MBC contribution, as shown in Fig.~\ref{fig5L} for several $g/J$. Interestingly, this MBC contribution is large when $\Delta$ is very far from the band-edges, differently from the 1D counterpart, such that its decay does not appear hidden by the BS contribution. Using a numerical fitting to a power law scaling, $1/t^\alpha$, for the range of times plotted in the figures one obtains $\alpha\approx 2.5$. However, a closer inspection to the integral dominating the long-
time dynamics of the MBC contribution leads to~\cite{wong78a}:
\begin{align}
\label{eqL:longpop}
\lim_{t\rightarrow\infty } C_{\mathrm{MBC}}(t)\propto \int_0^\infty \frac{e^{-y t}}{\log\left(\frac{y}{16J}\right)^2}\propto \frac{1}{t\log(t 16J)^2}\,,\,
\end{align}
such that the population, $|C_e(t)|^2$, in the long-time limit is $O[t^{-2}\log(t 16J)^{-4}]$. Notice that both Eqs.~\ref{eqL:singGam} and \ref{eqL:longpop} are directly connected to the logarithmic divergence of the density of states around $\Delta=0$ associated to the saddle point of $\omega(\kk)$. Thus, we conjecture that for other dispersion relations with saddle points, the same analysis will give rise to a similar quantitative behavior as that determined by those formulas.

Finally, it is instructive to see how the non-Markovian corrections affect to the population of the modes. Interestingly, in the long-time limit the shape of the outgoing wave-packet beyond the perturbative approach is still dominated by the pole occurring at $E=\omega(\kk)$:
\begin{align}
\label{eqL:ck}
 \lim_{t\rightarrow\infty}C_{\kk}(t)=\frac{g e^{-i\omega(\kk) t}}{\omega(\kk)-\Delta-\Sigma_e(\omega(\kk))}\,.
\end{align}
as the contributions from the UPs or BC will ultimately decay. As in the 1D case for, $g\ll J$ there is quasi energy conservation, that is, only $\kk$-modes satisfying $\omega(\kk)\approx \Delta$ are emitted. However, whereas in 1D this gives rise to a double Lorentzian structure of $C_k$ which peaks are centered at $k(\Delta)$ with width $\Gamma_e(\Delta)/2$, in 2D the situation is much richer, especially around $\Delta\approx 0$. In that case the emission occurs in modes with $\kk$ fulfilling $\omega(\kk)\approx0$, i.e., $ k_x\pm k_y\approx \pi (\mathrm{mod} 2\pi)$, which is why we see only emission in Fig.~\ref{fig2L} in real space around the diagonals $(n,\pm n)$.  Moreover, as $\mathrm{Re}(\Sigma_e)$ is discontinuous this translates into a double peak structure around the lines defined by $ k_x\pm k_y=\pi (\mathrm{mod} 2\pi)$, with a dip exactly at the lines which is strictly zero in the continuum limit as $\Gamma_e(0)\rightarrow \infty$.  The fact that we do not observe the dip in the population in Fig.
~\ref{fig2L} is attributed in this case to the finite size effects of the numerical simulation, which ultimately renormalizes $\Gamma_e$ to a finite value. However, it can be shown that for large enough systems (not shown), the $|C_\kk|^2$ indeed shows a dip at these $\kk$'s, which therefore agrees with the predictions of Eq.~\ref{eqL:ck}.

\begin{figure*}
\centering
\includegraphics[width=0.9\linewidth]{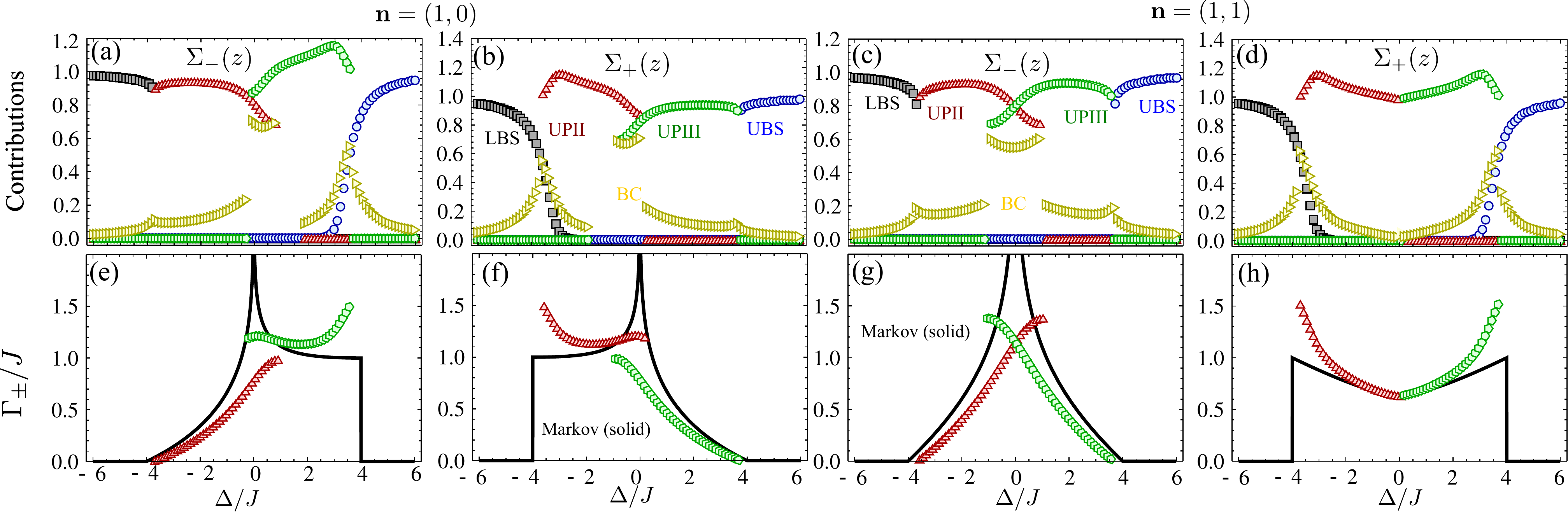}
\caption{(a-b) [(c-d)] Contributions to the probability amplitude $|C_\pm(t=0)|$  as a function of $\Delta$ for two QEs initially prepared in $\ket{\Phi_\pm}$ interacting with $g=J$ with relative position $\nn_{12}=(0,1)$ [$(1,1)$]. (e-f) [(g-f)] Imaginary part of the unstable poles for the symmetric/antisymmetric component $\Sigma_\pm$ for two QEs coupled with $g=J$ for $\nn_{12}=(0,1)$ [$(1,1)$] respectively.}
\label{fig6L}
\end{figure*}

\section{Two QE's dynamic for two-dimensional baths \label{sec:two}}

We consider now the situation with $N_e=2$ QEs to study the interplay between the relative position $\nn_{12}$ of the QEs and the directionality of the emission that we observed in the previous Section. In particular, we are interested in super/sub-radiant effects, which is why we study the problem of the QEs initially prepared in $\ket{\Phi_0}_S=\ket{\Phi_{\pm}}=\frac{1}{\sqrt{2}}(\sigma_{eg}^1\pm\sigma_{eg}^2)\ket{g,g}$.

As we explain in Section \ref{sec:system}, the probability amplitude $C_{\pm}(t)$ associated to the symmetric/antisymmetric superpositions can be calculated using Eq.~\ref{eqL:probabilityamp} for $C_e(t)$, but replacing the $\Sigma_e\rightarrow \Sigma_{\pm}=\Sigma_{e}\pm\Sigma_{12}$, where $\Sigma_{12}$ is given by Eq.~\ref{eqL:self12}. Before we give the general solution for $\Sigma_{12}(z,\nn_{12})$ for all $z$ and $\nn_{12}$, let us restrict our attention to QE energies close to the band edges  where $\omega(\kk)\approx \pm 4J \mp J k^2$ is isotropic. For example, close to the lower band-edge the collective self-energy is approximately given by:
\begin{align}
 \Sigma_{12}(-4J+E+i0^+;\nn_{12})\approx \frac{g^2}{4i J} H_0^{(1)}\left(\frac{|\nn_{12}|}{\sqrt{J/E}}\right)\,
\end{align}
where $H_0^{(1)}(x)=J_0(x)+i Y_0(x)$ is the Hankel Function of the first kind~\cite{abramowitz66a} and $\sqrt{J/E}$ is the associated wavelength of the excitations. This results agrees with the ones found in the literature using isotropic models \cite{douglas15a,gonzaleztudela15c}. 

In contrast, in this manuscript we are interested in studying the problem beyond the isotropic approximation. For that purpose it is enlightening to change the $\kk=(k_x,k_y)$ and $\nn_{12}=(n_x,n_y)$ variables in the integral to $k_{x,y}=q_x\pm q_y$ and $\bar{n}_{x,y}=n_x\pm n_y$, such that  in the continuum limit the integral transforms to
\begin{align}
\Sigma_{12}(z;\nn_{12})&=\frac{g^2}{\pi^2}\int_{0}^\pi\int_{0}^\pi d^2\qq\frac{\cos(q_x \bar{n}_x)\cos(q_y \bar{n}_y)}{z+4J\cos(q_x)\cos(q_y)}\,,
\end{align}
which now can be integrated by parts in certain regimes. For example, it is possible to show that~\cite{economoubook83a,morita71a}:
\begin{align}
 \label{eqL:greeen}
 \Sigma_{12}(z;(1,1))&=\frac{2 g^2}{\pi z}\left[\left(\frac{2}{m(z)}-1\right)K(m(z))-\frac{2}{m(z)} E(m(z))\right]\,,\\
 \Sigma_{12}(z,(1,0))&=\frac{1}{4J}\left[g^2-z\Sigma_e(z)\right]\,,
 \end{align}
where we define $m(z)=\left(\frac{4J}{z}\right)^2$ and where $E(m)=\int_0^{\pi/2}\sqrt{1-m\sin^2(x)} dx$ is the complete elliptical integral of the second kind. The possibility of obtaining such closed expressions for all $z$ allows us to make a separation of the different contributions to the probability amplitudes as we did for the single QE in Fig.~\ref{fig4L}(b). The result of this analysis for $\nn_{12}=(0,1)$ and $(1,1)$ is shown in Fig.~\ref{fig6L}(a-d). For simplicity, we have used a single color for the all the BC contributions, but as in the single QE situation we have an upper/lower/middle BC contribution. Let us highlight the main differences with respect to the single QE situation:

\emph{i)} Depending on $\nn_{12}$ LBS (or the UBS) associated with $\Sigma_{\pm}$ does not exist for all parameter regimes. For example, in Fig.~\ref{fig6L}(a-b), we see how for $\nn_{12}=(1,0)$ the UBS (LBS) of the $\Sigma_{+(-)}$ component disappears for a critical $\Delta$. For $\nn_{12}=(1,1)$ both BSs survive (disappear) for the $\Sigma_{+(-)}$ component. The reason for the disappearance of the BSs can be traced back to a cancellation of the divergence of $\delta\omega_e(E=-4J)$ by the collective component $J_{12}(E)$, as we already showed for 1D situation in Section~\ref{sec:1D}. 

\emph{ii)} More importantly for this manuscript is that something similar happens with the two UPs contributions appearing in the middle of the band. For example, for $\nn_{12}=(1,1)$ and $\Sigma_-(z)$ (see Fig.~\ref{fig6L}(c)] the region of coexistence of the two solutions enlarges by a factor 2 to $\Delta\in [-\frac{g^2}{J},\frac{g^2}{J}]$ with respect to the single QE situation, whereas in the symmetric one (see Fig.~\ref{fig6L}(d)) it disappears. This can be explained by the fact that the relative phases of $\Sigma_{12}$ and $\Sigma_e$ around that point make these functions add up constructively/destructively for the antisymmetric/symmetric components respectively. Moreover, in panels (g-h) we plot the imaginary part of the UP in the second and third Riemann sheet and show how these constructive/destructive interference in the antisymmetric/symmetric components corresponds to a super/subradiant behaviour manisfested in an enhanced/suppressed decay rate with respect to the single QE situation. Remarkably, 
the subradiant phenomena does not cancel completely the emission to the reservoir for 
two 
QEs, i.e., $\Gamma_{\pm}\neq 0$ in any combination $(d,\Delta)$, which is different from what happened with 1D reservoirs. We will give a more intuitive explanation for that behaviour below.

\begin{figure}
\centering
\includegraphics[width=0.98\linewidth]{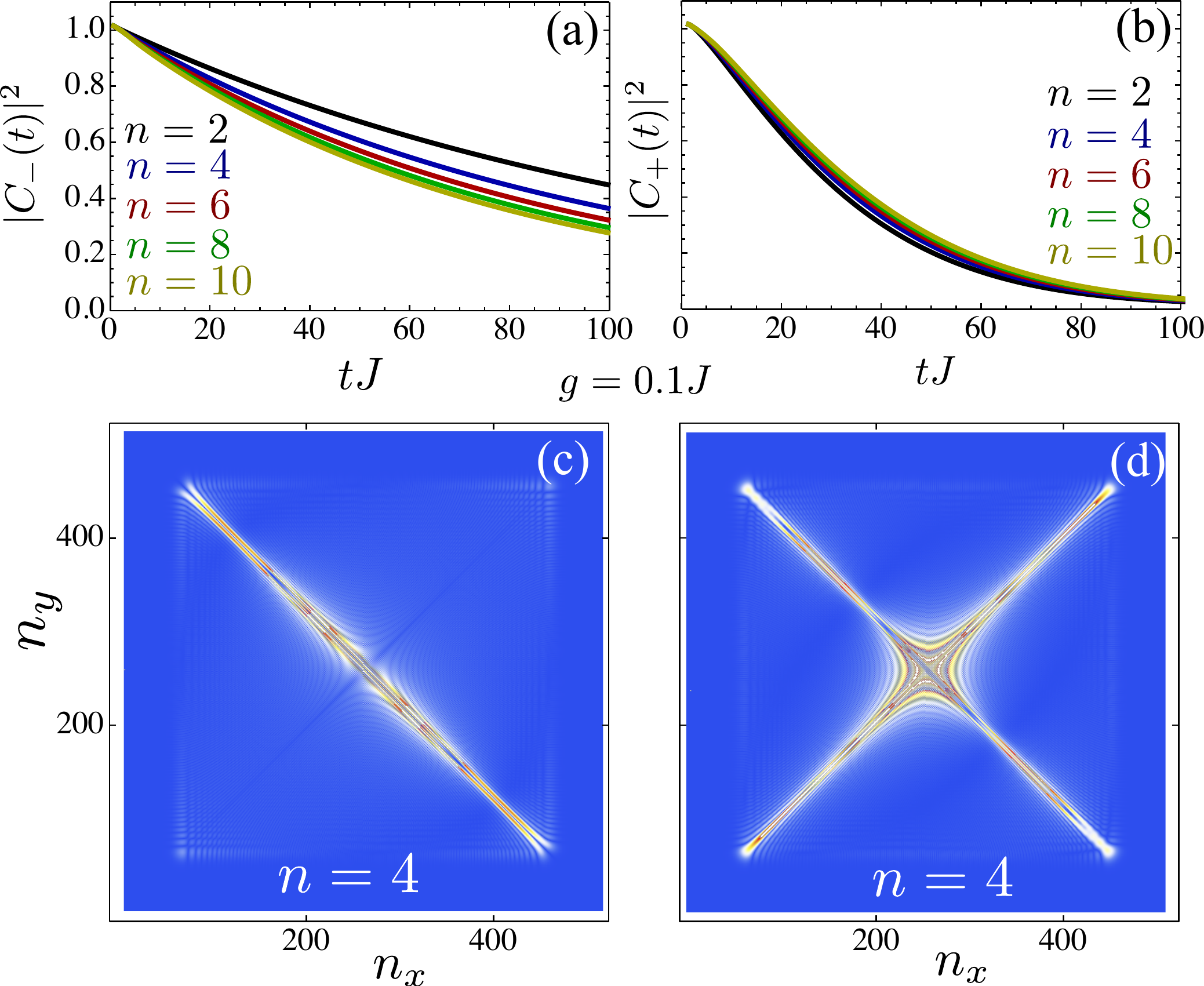}
\caption{(a-b) Symmetric/Antisymmetric population $|C_\pm(t)|^2$  for two QEs with $g=0.1J$ and $\Delta=0$ for different positions $\nn_{12}=(n,n)$ along the diagonal as depicted in the legend. (c-d) Corresponding bath population of the upper panels for $\nn_{12}=(4,4)$ at a time $tJ=100$.}
\label{fig7L}
\end{figure}

For the case $\nn_{12}=(1,0)$, the main difference is that it appears an asymmetry $\Delta\gtrless 0$. In particular, the coexistence of the two UPs occurs for a regime of $\Delta\in [0,\frac{g^2}{J}]$ ($[-\frac{g^2}{J},0]$) for the antisymmetric (symmetric) component. Then, looking at the imaginary components for the UPs, we observe that depending on whether $\Delta\gtrless 0$ either the symmetric/antisymmetric component behaves super/subradiantly. We have certified this type of behaviour for other positions (not shown), by using recursive relationships \cite{morita71a,economoubook83a} between the $\Sigma_{12}(z,\nn_{12})$ which allows one to obtain their analytical structure by starting with the ones of Eqs.~\ref{eqL:greeen}.

\begin{figure}
\centering
\includegraphics[width=0.8\linewidth]{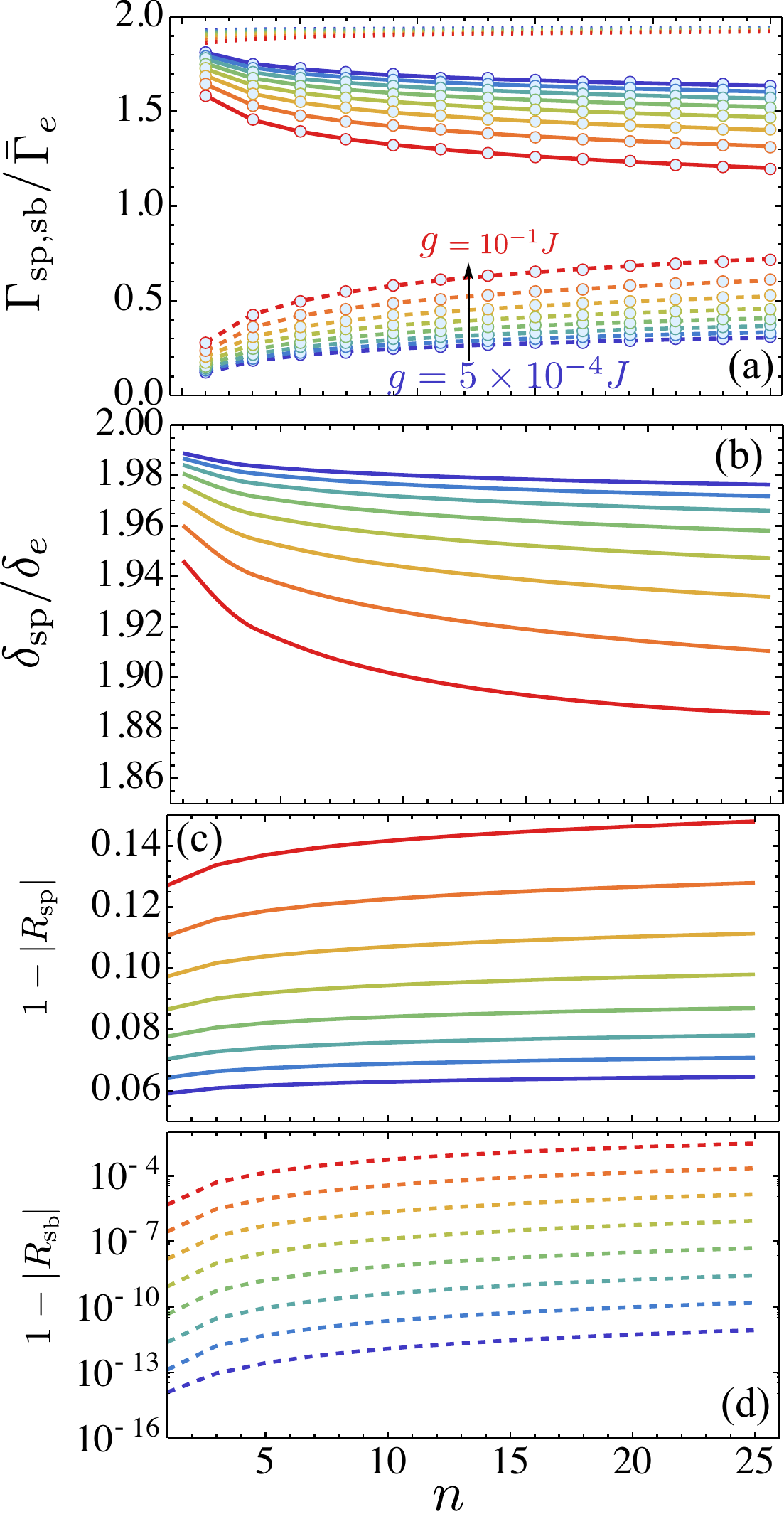}
\caption{(a) [(b)] Imaginary [Real] part of super/subradiant pole with solid/dashed lines compared to the one a single QE as a function of $\nn_{12}=(n,n)$ for 8 logarithmically spaced values of $g/J$ ranging from $5\times 10^{-4}$ to $10^{-1}$. In panel (a) in dotted lines we plot the sum of super/sub radiant states. In panel (b) we do not plot the one of the subradiant states as it is strictly $0$. (c-d) Corresponding residue of super/subradiant poles for the situation of panels (a-b)}
\label{fig8L}
\end{figure}

From the study of a single QE, we know that for $\Delta=0$ the emission occurs into two quasi-1D modes along the diagonal. This is why, from now on, we focus on the situation where the two QEs are placed along a diagonal, e.g., $\nn_{12}=(n,n)$, for which the recursive relation simplifies to:
\begin{align}
 \Sigma_{12}(z;(n+1,n+1))&=\frac{4 n}{2n+1}\left[\frac{2}{m(z)}-1\right]\Sigma_{12}(z;(n,n))-\nonumber\\
 & \frac{2n-1}{2n+1}\Sigma_{12}(z;(n-1,n-1))\,.
\end{align}

From this formula it can be obtained that when the QEs lie in the middle of the band ($\Delta=0$),  it seems that the Markov description, the $\Sigma_{12}(i 0^+;(n,n))/\Sigma_{e}(i 0^+)=(-1)^n$, which points to two interesting conclusions: i) first, the alternating phase $(-1)^n$ tell us that the symmetric/antisymmetric state will decay super(sub) radiantly for even (odd) $n$. ; ii) On the other hand, the fact that they have the same amplitude points to the possibility of observing perfect super/subradiant behaviour as in 1D systems. However, as it occurs for the single QE both $\Sigma_{e},\Sigma_{12}$ diverge at this point, such that one has to be cautious when applying Markovian approximation~\footnote{The claim (ii) seems to disagree with the numerical results of Ref.~\cite{galve17a}, where the quotient $\Sigma_{12}(i 0^+;(n,n))/\Sigma_{e}(i 0^+)=(-1)^n/2$ for large distances $n$. Even though the model is different from ours, we believe that the discrepancy may come from the numerical evaluation of 
integrals of $\Sigma$'s, which in their case are time dependent. }.

In order to understand how Markovian approximation gets corrected in this situation we numerically integrate the exact dynamics of the (anti)symmetric state for different even $n=2,4,\dots 10$ for a situation with $g=0.1J$, in which the QEs should decay sub(super)radiantly. We show the corresponding dynamics in Fig.~\ref{fig7L}(a-b) and complement in panels (c-d) with the bath population of the $n=4$ situation at $t J=100$. Interestingly, the (almost) perfect superradiant behaviour is observed in which the decay rate seems to vary very weakly with the QE positions. On the other hand, as expected, the perfect subradiant behaviour is not obtained.

An intuitive picture for the absence of perfect subradiance can be obtained by looking at the bath population dynamics, where we see that the subradiant state is able to cancel bath emission in the diagonal where the QEs are placed, but the QEs are free to decay in other one as shown in Fig.~\ref{fig7L}(c). The cancellation along the QE diagonal occurs because the bath excitations emitted in both QEs add up 
destructively, which ultimately traps the excitation as we show it occurs in 1D setups (see Section~\ref{sec:1D}). However, the other diagonal creates an open decay channel that forbids the existence of perfect subradiant behaviour. In Section \ref{sec:many}, we will see how to add more QEs to recover the perfect subradiance of 1D systems.

In order to get a better understanding on the scaling of the super/subradiance phenomena for two QEs with the distance and the relative value $g/J$, we solve exactly the pole equation for $\Delta=0$, i.e., $z=\Sigma_{\pm}(z;(n,n))$, to find the corresponding UPs in the second/third Riemann sheet. In order to do it, we have to use the analytical continuation of $K(m)$ and $E(m)$ in those sheets. For $K(m)$ we already give them in Eq.~\ref{eqL:selfIIandIII} for the single QE situation. For the elliptic integral of the second kind, they read~\cite{abramowitz66a,morita71a} $E^{\mathrm{II,(III)}}(m)=E(m)\pm 2i\left[\KK(1-m)-E(1-m)\right]$.

The results are summarized in Fig.~\ref{fig8L}(a-b), where we plot the imaginary and real part of the UPs (normalized by the individual decay rate) corresponding to a superradiant (solid) and subradiant configuration (dashed) as a function of $n$ for different $g/J$ ranging from $5\times 10^{-4}$ (blue) to $10^{-1}$ (red). In dotted lines we also plot the sum of the super and subradiant decay rates $\left(\Gamma_\mathrm{sp}+\Gamma_{\mathrm{sb}}\right)/\bar{\Gamma}_e$. The super(sub)radiant states initially have enhanced (suppressed) decay rate close to $2 (0)\bar{\Gamma}_e$ converging very slowly to $\bar{\Gamma}_e$ as the distance increases as ultimately, the QEs decay with the same timescale than they would have individually. The smaller the ratio $g/J$ is, the closest is the initial value to the ideal situation and the slower is the convergence to the individual situation. Another remarkable consequence is that $(\Gamma_\mathrm{sb}+\Gamma_\mathrm{sp})/\bar{\Gamma}_e<2$ (plotted in dashed lines in Fig.~\ref{fig8L}(a) for short distances, breaking the sum rule typical of Markovian evolution. 

The real part of the subradiant state is exactly $0$, as we already showed in Figs.~\ref{fig6L} that the subradiant configuration is characterized by canceling the discontinuity of the real part at $\Delta=0$. On the contrary, the real part of $\Sigma_e$ and $\Sigma_{12}$ add up constructively for the superradiant configuration, which is why the real part of the UP is very close to $2\delta_e$ as shown in Fig.~\ref{fig8L}(b). Remarkably, it also shows a very slow decay, being almost collective for the plotted range of distances.  This provides an exciting perspective in which both long-range coherent interactions and collective decay can be obtained, which is different from the 1D scenario in which only one of them survives.

For $\Delta=0$ the dynamics is dominated mainly by the contribution of the UPs and the MBC~[see Fig.~\ref{fig6L}]. The former gives a contribution to the dynamic $|C_\pm(t)|^2\propto R_{\mathrm{sp},\mathrm{sb}} e^{-\Gamma_{\mathrm{sp},\mathrm{sb}}t}$, where $R_{\mathrm{sp},\mathrm{sb}}$ is the associated residue, whereas the MBC gives rise to subexponential decays. In Fig.~\ref{fig8L}(c-d), we plot the deviation from $1$ of the associated residue of the UPS to see how important are the corrections to the Markovian predictions.  There, we find that, i) the non-Markovian corrections increase with the distance $n$, because of retardation effects; ii) but also that the superradiant configuration has a larger correction than the subradiant one, because of the coexistence of two UPs.

\begin{figure}
\centering
\includegraphics[width=0.8\linewidth]{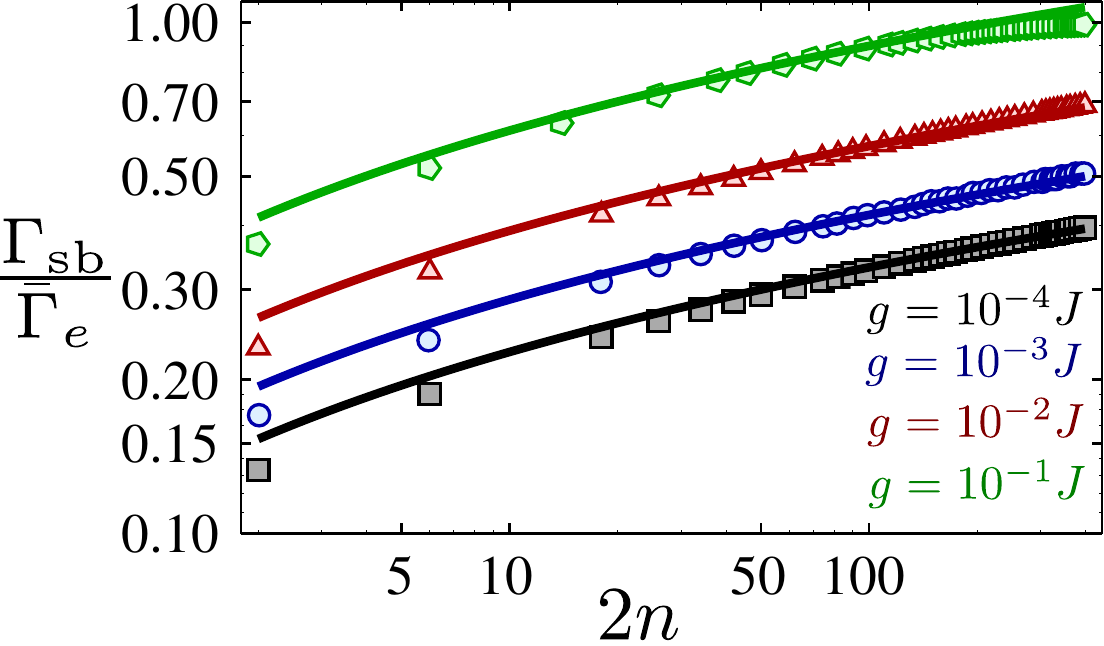}
\caption{Imaginary part of subradiant pole of two QEs with positions $\nn_{12}=(2n,2n)$ for  $g/J=10^{-4},10^{-3},10^{-2},10^{-1}$. Markers correspond to the numerical solution of the pole equation, whereas the solid lines correspond to the asymptotic expansion of Eq.~\ref{eqL:asyn}.}
\label{fig8Lbis}
\end{figure}

Finally, to get a better quantitative understanding on the dependence with the distance, we write explicitly the pole equation for the subradiant state occurring for $\nn_{12}=(2n,2n)$, which correspond to the antisymmetric configuration:
\begin{align}
 \label{eqL:sub}
 \bar{z}_\mathrm{sb}&=8\left(\frac{g}{8\pi J}\right)^2\iint_{0}^\pi d\qq \frac{\sin^2(2 q_x n)}{\bar{z}_\mathrm{sb}+\cos(q_x)\cos(q_y)}
\end{align}
where $\bar{z}=z/(4J)$. As we know from the numerical results of Fig.~\ref{fig8L}, that the real part of the subradiant pole is strictly zero, we can write directly the pole equation for the imaginary component $\bar{z}=i y$, and perform the $q_y$ integral to arrive to:
\begin{align}
 \label{eqL:sub2}
 8\pi\left(\frac{J}{g}\right)^2&=\int_{0}^\pi dq_x \frac{\sin^2(2 q_x n)}{y\sqrt{y^2+\cos^2(q_x)}}
\end{align}
where we have chosen negative branch of the $\sqrt{.}$ as we are looking for the solutions of the analytical continuation of $G_{-}(z)$ in the lower half plane. Eq.~\ref{eqL:sub2} is much easier to solve, which allows us to go to much larger distances as shown in Fig.~\ref{fig8Lbis}. Interestingly, integrating by parts and making an expansion first for $y\ll 1$ and then for $n\gg1$, the subradiant decay can be asymptotically approximated by:
\begin{equation}
\label{eqL:asyn}
 \Gamma_{\mathrm{sb}}\approx \frac{g^2}{\pi J}\left(\gamma+\log(8n)\right)\,.
\end{equation}
where $\gamma\approx 0.577$ is the Euler constant. We compare this formula with the exact solution in Fig.~\ref{fig8Lbis} showing an excellent agreement for $g\ll J$ and $n\gg 1$. Notice that Eq.~\ref{eqL:asyn} is only valid for $\Gamma_\mathrm{sb}/J\ll 1$, and obviously gets corrected for larger distances where $\Gamma_\mathrm{sb}\approx \bar{\Gamma}_e$ as one expects to recover the independent emission result. Remarkably, the two QE arrive to the situation of independent emission with a very weak dependence with the distance ($\log(n)$).

\section{Several QE's dynamic for two-dimensional baths \label{sec:many}}

After having explored extensively the dynamics of a single and two QEs, we finally consider the dynamics of several QEs. As before, we restrict to the single excitation dynamics and for the sake of concreteness we avoid the discussion on the different contributions to the dynamics obtained from the analysis of the complex integration. Here, we adopt a pragmatic perspective in which we use the intuition developed in the previous Sections to search novel effects. 

In particular we want to answer two questions: i) is it possible to find perfect subradiant states in our setup?, ii) how does the superradiance behave when increasing the number of QEs? We answer these questions in Section \ref{sec:four} and \ref{sec:super} respectively.

\subsection{Four QEs: Square-like subradiant states \label{sec:four}}

\begin{figure}
\centering
\includegraphics[width=0.8\linewidth]{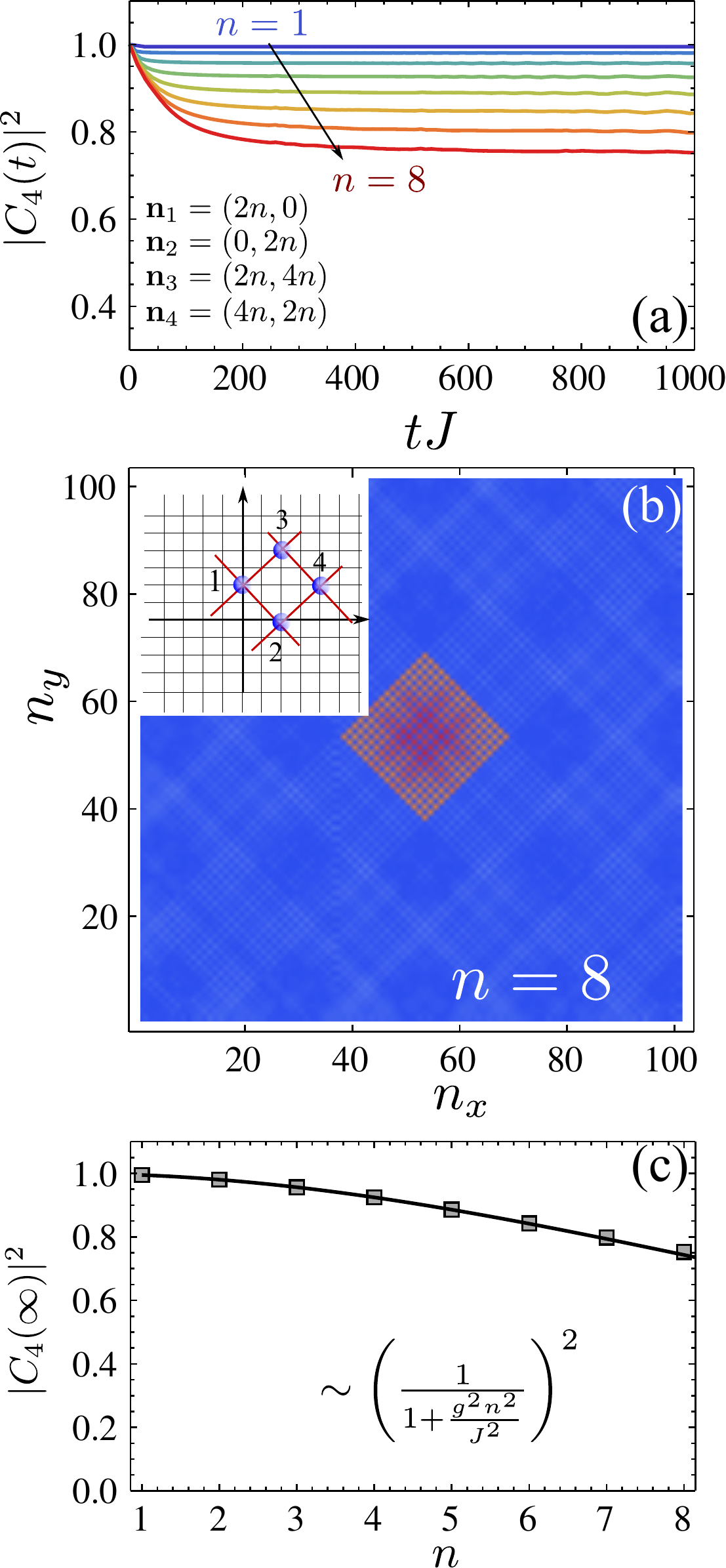}
\caption{(a) Population dynamics $|C_4(t)|^2$ for four QEs with $\Delta=0$, $g=0.05J$ and positions as depicted in the legend. (b) Bath population in real space for a time $tJ=1000$, where we observed the population trapping in a square given by the positions of the QEs for the value of $n=8$ in panel (a). Inset: Scheme of the cancellation in the 8 directions where the QEs are emitting into the bath. (c) $|C_4(\infty)|^2$ as a function of $n$ for $g/J=0.05$ numerically obtained (markers) compared to analytical expression obtained in the text.}
\label{fig9L}
\end{figure}

Analyzing the propagation of the bath excitations, we show in Section \ref{sec:two} that by placing two QEs at $(0,0)$ and $(2n,2n)$ prepared in an antisymmetric superposition $\ket{\Psi_-}$, the population decay gets reduced because of the destructive interference along the diagonal where the QEs are present. However, the emission into the orthogonal direction provides a decay channel that can not be suppressed with only two QEs. In this Section we show how by placing four QEs appropriately a perfect cancellation of spontaneous emission into the bath can be obtained.

The Hamiltonian interaction for four QEs is given by $H_\mathrm{int}$ of Eq.~\ref{eqL:Hintt} with $N_e=4$. However, to understand the effect it is instructive to rewrite the spin operators in the following four orthogonal states:
\begin{align}
 \sigma^{a,b}_{eg}&=\frac{1}{2}\left(\sigma_{eg}^1\pm \sigma_{eg}^2+\sigma_{eg}^3\pm\sigma_{eg}^4\right)\,,\\
 \sigma^{c,d}_{eg}&=\frac{1}{\sqrt{2}}\left(\sigma_{eg}^{1,2}-\sigma_{eg}^{3,4}\right)\,.
\end{align}

Using those operators, the interaction Hamiltonian reads:
\begin{align}
 H_\mathrm{int}=\frac{g}{N}\sum_{\kk,\alpha=a,b,c,d}\left( f_\alpha(\kk) a_\kk \sigma_{eg}^\alpha +\mathrm{h.c.}\right)\,,
\end{align}
where $f_\alpha(\kk)$ are the mode functions coupling to the $\alpha$-mode, which are given by:
\begin{align}
 \label{eqL:mode}
 f_{a,b}(\kk)&=\frac{1}{2}\left(e^{i \kk\cdot \nn_1}\pm e^{i \kk\cdot \nn_2}+e^{i \kk\cdot \nn_3}\pm e^{i \kk\cdot \nn_4}\right)\,,\\
  f_{c,d}(\kk)&=\frac{1}{\sqrt{2}}\left(e^{i \kk\cdot \nn_{1,2}}\pm e^{i \kk\cdot \nn_{3,4}}\right)\,,
  \end{align}

  In general, the dynamics of the four QEs is complicated because all the modes are coupled between themselves as they couple to non-orthogonal modes of the bath, i.e., $f_\alpha(\kk)a_\kk$. However, by fixing the positions $\nn_{1}=(2n,0),\nn_{2}=(0,2n),\nn_{3}=(2n,4n)$, $\nn_4=(4n,2n)$, we can prove how the dynamics can be simplified. To show it, we change the $\kk$ variables in $H_\intt$ to $k_{x,y}=q_x\pm q_y$, and rewrite the $\qq$ sum restricting to the ones with $q_{x,y}>0$, we arrive to
  \begin{align}
   \label{eqL:Hint}
   H_\intt=\frac{g}{N}\sum_{\qq>0,\alpha} N_{\alpha}(\qq,n)\left(a_{\alpha,\qq}\sigma_{eg}^\alpha+\mathrm{h.c.}\right)\,.
  \end{align}

Now, the $a_{\alpha,\qq}$ are the normalized bath modes that couple to the $\sigma_{eg}^\alpha$ QE state, with a mode function $N_{\qq,\alpha}$ which can be obtained from the normalization condition. For example, the $\alpha=a,b$ modes are defined
  \begin{align}
   \label{eqL:orth}
   a_{a/b,\qq}&=\frac{1}{N_{a/b}(\qq)}\sum_{\theta=1}^4 f_{a/b}(\hat{R}_{\theta}\qq) a_{\hat{R}_{\theta}\qq}\,\,,\\
   N_{a,b}(\qq,n)&=2\sqrt{1+\cos(4 q_x n)\cos(4 q_y n)\pm\cos(4 q_x n)\pm\cos(4 q_y n)}
  \end{align}
where $\hat{R}_{\theta}$ denotes the rotation of the $\qq$-variables an $\theta\pi/2$ angle. Interestingly, these modes can be shown to be orthogonal to the rest, which means that their dynamics decouple and can be treated independently. Moreover, from the intuition developed in the previous Sections, we know that by fixing $\Delta=0$ the bath modes dominating the dynamics are such that $k_x+k_y=2q_x=\pm \pi$, which can be shown to lead to $N_{a}(\qq,n)=4$ and $N_{b}(\qq,n)=0$ independent of the distance which points to both having a perfect super and subradiant effect in the dynamics. This effect can be interpreted from the emission in those $\kk$, which will propagate for each QE in two directions, interfering constructively/destructively along the total 8 spatial directions as sketched in Fig.~\ref{fig9L}(b)).

Now, let us explore more rigorously the subradiant behaviour and study the dynamics of the four QEs when they are initialized $\ket{\Phi_4}=\sigma_{eg}^b\ket{g}^{\otimes 4}$. As we have just seen that its dynamics decouples from the rest of the modes, we can apply the resolvent operator technique to calculate the probability amplitude $C_4(t)$. In particular, $C_4(t)$ can be obtained from the same Fourier transform than the one of a single QE, but replacing the individual self-energy $\Sigma_e(z)$ by:
\begin{align}
 &\Sigma_{4}(z)=\frac{g^2}{4\pi^2}\iint_{0}^\pi d\qq\frac{|N_{b}(\qq,n)|^2}{z+4J\cos(q_x)\cos(q_y)}\,,
\end{align}
where $|N_{b}(\qq,n)|^2=4\big[1+\cos(4 q_x n)\cos(4 q_y n)-\cos(4 q_x n)-\cos(4 q_y n)\big]=16\sin^2(2nq_x)\sin^2(2nq_y)$. As we did for the 1D case in Section~\ref{sec:1D} to prove that the state $\ket{\Phi_4}$ is perfectly subradiant it suffices to show: i) $\Sigma_4(0)=0$, that will show that the imaginary part of the UPs is zero at $\Delta=0$; and ii) that the associated residue, $R_4$, is finite. The later is important because it directly relates to the remaining excitation in the QE state in the final steady state, i.e., $C_4(\infty)=R_4$. We start by proving i), by showing that $\Sigma_4(0)$ is given by:
\begin{align}
 &\Sigma_{4}(0)=\frac{g^2}{\pi^2 J}
 \iint_{0}^\pi d\qq\frac{\sin^2(2q_x n)\sin^2(2q_y n)}{\cos(q_x)\cos(q_y)}\,,
\end{align}
that is, a separable integral in $q_{x,y}$. There, one can show that the numerator [denominator] of the integrand, $\sin^2(2 n q)$ [$\cos(q)$], is an even [odd] function with respect to $\pi/2$ in the interval of integration $[0,\pi]$. Moreover, the divergence at $q_x=q_y=\pi/2$ of the denominator is cancelled because the numerator goes to $0$ as well faster than the denominator. Thus, $\Sigma_4(0)\equiv 0$. The only remaining point is to calculate the residue using that $R_4=\left(1-\partial_z \Sigma_4(z)\right)^{-1}|_{z=0}$. It can be shown that the only non-zero contribution is
\begin{align}
 \label{eq}
  \partial_z\Sigma_4(z)|_{z=0}=-\frac{g^2}{4\pi^2 J^2}\iint_{0}^\pi d\qq \frac{\sin^2(2nq_x)\sin^2(2nq_y)}{\cos^2(q_x)\cos^2(q_y)}\,,
\end{align}
where:
\begin{equation}
 \int_{0}^\pi dq \frac{\sin^2{2nq}}{\cos^2(q)}=2\pi n\,.
\end{equation}

Thus, the final steady-state population remaining in $\ket{\Phi_{4}}$ is given by:
\begin{align}
\label{eqL:R4}
 C_4(\infty)=\frac{1}{1+\frac{g^2 n^2}{J^2}}\,.
\end{align}

As in the 1D situation, the perfect interference appears when there is no which-way information on which QE emitted the photons. In 2D, the correction to the perfect interference is given in Eq.~\ref{eqL:R4} by the term $g^2 n^2/J^2 $, which contains both the ratio between the timescales required for the QEs to decay to the bath compared to the propagation of excitations in it ($g^2/J^2$), and the $n^2$ factor, different from the 1D counterpart ($n$), which we attribute to the fact that the interference must occur in two directions simultaneously. We emphasize as we did for the 1D situation, that the state $\ket{\Phi_4}$ remains perfectly subradiant in all the parameter regimes, but only its overlap with the initial state decreases.

We numerically certify these predictions and summarize the results in Fig.~\ref{fig9L}. In Fig.~\ref{fig9L}(a), we plot the dynamics of $|C_4(t)|^2$ for four QEs coupled with $g=0.05J$ and for different distances controlled by the value $n$ of the vector positions $\nn_i$. There, we observe that indeed the spontaneous decay into the bath gets suppressed compared to the situation of two QEs. We complement these figure by plotting the population in the bath modes in real space in Fig.~\ref{fig9L}(b) for a particular separation, i.e., $n=8$, where we observe the population mostly remains trapped in the square defined by the positions of the QEs. Obviously, for larger distances retardation effects enters into play, as the excitations needs to propagate from one QE to the other before the destructive interference can play a role, as it occurs for the waveguide QED. Finally, in Fig.~\ref{fig9L}(c), we compare the steady state numerical results for $|C_4(\infty)|^2$ with the 
prediction obtained in Eq.~\ref{eqL:R4}, $|R_{4}|^2$ showing how both the analytical and numerical results match very well.

\begin{figure}[tb]
\centering
\includegraphics[width=0.8\linewidth]{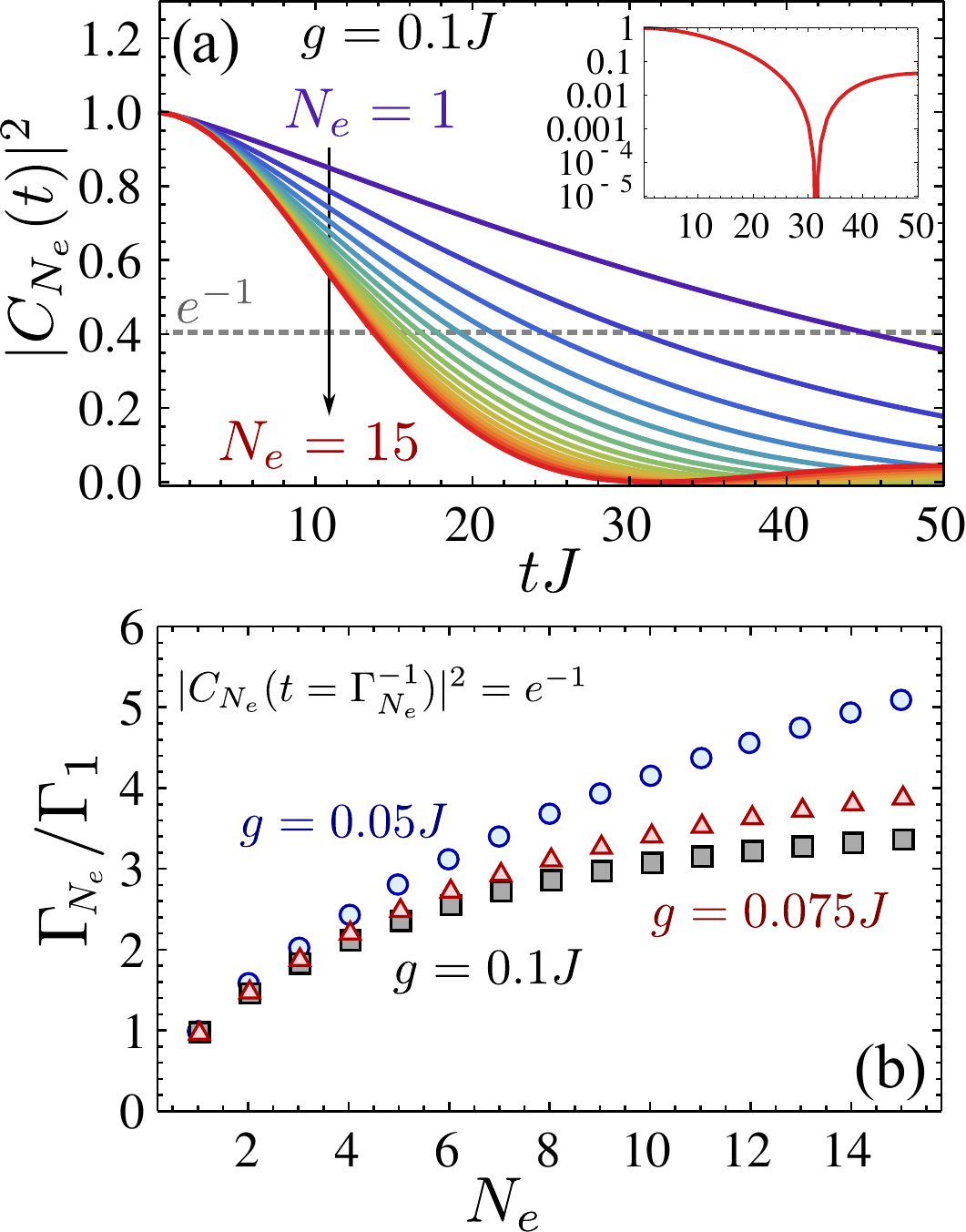}
\caption{(a) $|C_{N_e}(t)|^2$ for $g=0.1J$ for different $N_e=[1-15]$ as depicted in the legend at positions $\nn_j=(2j,2j)$. $j=0,\dots, N_e-1$. Inset: $|C_{N_e}(t)|^2$ in logaritmic scale where the non-Markovian effects appear more clearly. (b) $\Gamma_{N_e}/\Gamma_1$ for $g/J=0.05,0.075$ and $0.1$ obtained by solving the equation: $|C_{N_e}(t=1/\Gamma_{N_e})|^2=e^{-1}$.}
\label{fig10L}
\end{figure}

\subsection{Many QE: collective emission \label{sec:super}}

In Section~\ref{sec:two} we show how the two QE superradiant state experiences both long-range coherent interaction and collective decay which decreases very slowly with the distance [see Fig.~\ref{fig8L}]. It is very easy to show that adding new emitters will keep the interference.  Here we will concentrate in superradiance in the presence of several QEs when prepared initially prepared in a state:
\begin{equation}
 \label{eqL:sym}
 \ket{\Phi_{N_e}}=\frac{1}{\sqrt{N_{e}}}\sum_{j=1}^{N_e}\sigma_{eg}^j\ket{g}^{\otimes N_e}
\end{equation}
placed at positions $\nn_j=(2j,2j)$ to enforce that the decay collectively. The results are shown in Fig.~\ref{fig10L}(a-b). Let us highlight the most remarkable features: In Fig.~\ref{fig10L}(a), we indeed observe that the emission gets faster the more emitters we include in the calculation for a situation with $g=0.1J$. However, there is a point in which the increase in the decay rate is very slow. We quantify this observation by assuming an exponential law for the population $|C_{N_e}(t)|^2$ and defining $\Gamma_{N_e}$ as the inverse of time when $|C_{N_e}(t)|^2$ has decayed to $e^{-1}$. We plot the resulting scaling of $\Gamma_{N_e}$ in Fig.~\ref{fig10L}(b), together with the results for $g/J=0.05, 0.075$. We emphasize that this is only an approximation, as the dynamics is not exponential in the short/long time limit.  The short time corrections, $1-|C_{N_e}(t)|^2\propto t^2$, are a general feature of the decay to any reservoir \cite{cohenbook92a}, whereas the overdamped oscillations that we observe in 
the 
inset of Fig.~\ref{fig10L}(a) are attributed the coexistence of two complex poles with different real 
component, as we already explain in Section \ref{sec:two}. 

We attribute the saturation of $\Gamma_{N_e}$ to the logarithmic decay with the distance that we show in Figs.~\ref{fig8L}~\&~\ref{fig8Lbis} which introduces a finite length scale in which the constructive interference can be built. As we already show in those figures, the smaller the ratio $g/J$, the larger the length scale of the interactions which is why in that limit the saturation of $\Gamma_{N_e}$ occurs at a larger value in Fig.~\ref{fig10L}(b).

\section{Experimental considerations \label{sec:experimental}}

As we mentioned in Section~\ref{sec:system}, there are several platforms where the Hamiltonian $H=H_S+H_B+H_\intt$ can be implemented. In this Section, we discuss more in detail two of these implementations, namely, cold atoms in state dependent optical lattices~\cite{devega08a,navarretebenlloch11a} and quantum emitters coupled to engineered dielectrics~\cite{vetsch10a,huck11a,hausmann12a,laucht12a,thompson13a,goban13a,lodahl15a,sipahigi16a}. In particular, we analyze the conditions under which the predicted phenomena could be observed

\subsection{State dependent optical lattices}

\begin{figure}[tb]
\centering
\includegraphics[width=0.7\linewidth]{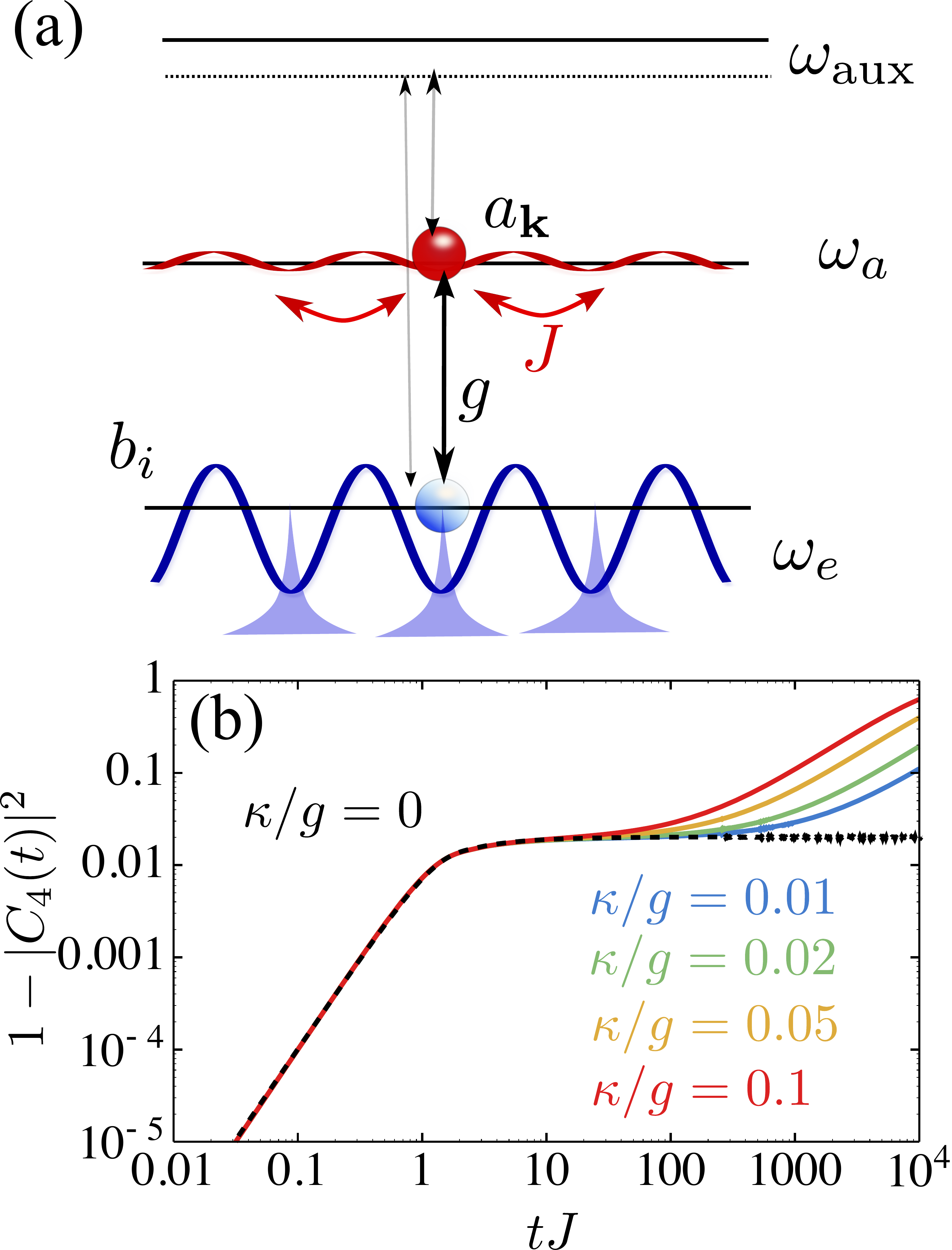}
\caption{(a) Scheme of a state-dependent optical lattice to simulate QED Hamiltonians. (b) Impact of bath loss, $\kappa$, on the dynamics of the perfect subradiant state of four emitters. The positions of the QE are as in Fig.~\ref{fig9L} with $n=1$. The rest of the parameters are $g=0.1J$, $\Delta=0$ and several $\kappa$'s as depicted in the legend.}
\label{fig11L}
\end{figure}

The QED Hamiltonian $H$ can be obtained in a state-dependent optical lattice with a setup as depicted in Fig.~\ref{fig11L} (a)~[see  Refs.~\cite{devega08a,navarretebenlloch11a} for an extended discussion]: a bosonic atom with two internal metastable states, that we label as $a$ and $b$, is trapped within an optical lattice such that the trapping potential suffered by both metastable states is very different. One internal state, e.g., $b$, with energy $\omega_e$, sees a deep potential such that the Wannier functions associated to each site $\nn$ are strongly localized. Thus,  they do not overlap between them, and therefore do not hop from one site to the other. Furthermore, since we will have just one atom, we can ignore all interactions. The Hamiltonian describing the dynamics of such internal state is then given by:
\begin{equation}
H_b\approx \sum_\nn \omega_e b^\dagger_\nn b_\nn\,.
\end{equation}

Since we have a single atom, we can write $b_\nn$ as an effective two-level system, i.e.,  $b_\nn\rightarrow \sigma^\nn_{ge}$  such that $H_b\approx H_S$. Thus, the $b$-atoms will play the role of the QEs.

The other internal state, $a$, with energy $\omega_a$ is trapped by a much weaker potential which makes the wavefunction less localized, therefore overlapping between nearest neighbours and allowing them to hop from one site to another. The Hamiltonian describing the dynamics of this internal state is given by:
\begin{equation}
H_a\approx \sum_\nn \omega_a a^\dagger_\nn a_\nn- J \sum_{\mean{\nn,\mm}}a^\dagger_\nn a_\mm\,,
\end{equation}
 where $J$ is the nearest-neighbour tunnelling rate. We can immediately identify $H_a\approx H_B$, that is, with the bath Hamiltonian we used along the manuscript. Both of them are trapped within the Lamb-Dicke limit, such that they are in their motional ground state. The only element left is to implement the coupling between the $a$-$b$ atoms. In the case where $a$ and $b$ are two mestastable states of an Alkaline atom the coupling can be obtained through a two-photon Raman transition using an auxiliary excited state as sketched in Fig.~\ref{fig11L}(a). Alternatively, Alkaline-Earth atoms have a mestastable optically excited state (that can survive for extremely long times) which allows one to use a direct optical transition between $a$ and $b$. The latter has the advantage that the energy difference between $\omega_a$ and $\omega_e$ lies in the optical regime which allows one to create very different potentials for the two internal states, avoiding scattering events~\cite{bloch08a}.
 
 However, both the finite lifetime of the excited state in the case of Alkaline-Earth atoms, or the scattering losses introduced by the trapping potential induce decoherence in both the QE and bath-like modes. These decoherence rates in realistic implementations can be as small as $\kappa, \Gamma^*\sim$ Hz~\cite{bloch08a,schreiber15a,daley08a}. As typical tunnelling rates are of the order of $J\sim 10$ KHz~\cite{bloch08a} and the coupling can be comparable to them, $g\sim O(J)$, then  $\kappa,\Gamma^*\lesssim 10^{-4}g$. 
 
 Assuming that $\kappa=\Gamma^*=\Gamma_\mathrm{loss}$ and using the solutions that we obtained in Section~\ref{subsec:loss}, we can estimate the conditions to observe the phenomena that we predicted in this implementation. For example, to observe the exponential decay and directionality for a single QE we require $\bar{\Gamma}_e/\Gamma_\mathrm{loss}\gg 1$, which is favoured by the logarithmic enhancement that occurs in the middle of the band. The oscillatory non-Markovian behaviour will be more challenging to observe as their timescale do not benefit from the logarithmic enhancement, such that the condition to be observed is $g^2/\left(J \Gamma_\mathrm{loss}\right)$, which is still within the reach in the state-dependent cold atom setup.  To observe superradiant behaviour with $N$ QEs, the former condition benefits in the weak coupling regime from an extra factor $N$, i.e., $N\bar{\Gamma}_e/\Gamma_\mathrm{loss}\gg 1$. Finally, the subradiant mode with $N=4$ QEs acquires a finite lifetime given by $\Gamma_\mathrm{loss}$, such that subradiance will only be observed within a timescale, $t\ll \Gamma_\mathrm{loss}^{-1}$.  
 
\subsection{QEs coupled to engineered dielectrics}

Another platform to observe the predicted phenomenology is the one consisting on QEs coupled to engineered dielectrics. In these setups, the bath modes are guided photonic modes confined into the dielectric structures, which can be patterned to display similar energy dispersion relations displaying the same features as the ones predicted by our model. The two-level system of the QEs can be an optical transition of either natural or solid state atoms. Alternatively, to increase the tunability, one can use a $\Lambda$ transition naturally appearing in optical QEs, where one metastable state, $g$, is driven by an off-resonant Raman laser to an optically excited state, $s$, that can decay to the guided photons in the dielectric into another metastable state, $e$. If the detuning is large enough, the excited state $s$ is only virtually populated and the dynamics of the metastable states, $g,e$, mimics the one of $H$. The consequence of the adiabatic elimination of the excited state is that both the spontaneous 
emission, of the order of MHz for Alkali atoms, and coupling constant are renormalized as $\Gamma^* x^2$ and $g\rightarrow g x$, being $x$ a small tuneable parameter depending on the Raman laser intensity and detuning. Like this, one can go to a limit where the dominant decay rate is the one given by $\kappa$. In Ref.~\cite{douglas15a}, the predicted coupling constant $g\sim 10 $ GHz, such that $\kappa/g\sim 10^{-2} $ for state-of-the-art Q factors.

In general, the conditions to observe the phenomenology will be similar to the ones described in the previous Section. However, in the subradiant case they became more favourable as the quantum jumps only occur in the bath modes which are, in that case, weakly populated in the non-retarded regime. Thus, the timescale to observe the subradiant plateau $t\ll \left[\left(1-|C_4(\infty)|^2\right)\kappa \right]^{-1}$, where the factor $\left(1-|C_4(\infty)|^2\right)$ takes into account the bath population trapped within the QEs. For illustration, we numerically calculate the impact of bath loss on the perfect subradiant states of Fig.~\ref{fig9L}. We consider a situation with $\Delta=0$, $g=0.1J$, fixed distance between QEs $n=1$, and plot $|C_4(t)|^2$ for several $\kappa$'s. As expected, the subradiant state is still formed but it acquires now a finite lifetime as predicted by Eqs.~\ref{eqL:solution} in Section~\ref{subsec:loss}.

\section{Conclusions and outlook \label{sec:conclu}}

Summing up, in both the main text \cite{gonzaleztudela17b} and along this manuscript we have shown both analytically and numerically how the coupling to a \emph{structured} 2D reservoir with square like geometry gives rise to a zoo of phenomenology with no analogue in other type of reservoirs. In particular,
\begin{itemize}
 \item For a single QE we show that by tuning the energy of the QEs one can explore very different regimes. For energies inside the band but close to the band edges, the single QE experiences a standard exponential relaxation with a decay rate predicted by Fermi's Golden rule and with an associated isotropic emission into the bath. As the energies get closer to the middle of the band, the emission gets more anisotropic until it emits into two quasi-1D orthogonal modes for $\Delta=0$~\cite{mekis99a,langley96a}. Moreover, the QE relaxation dynamics at this point is non-conventional as for short times is approximately exponential but with a timescale different than the one predicted by Fermi Golden rule, whereas for longer times is accompanied by an oscillation and a subexponential decay.
 
 We emphasize that the directionality comes from the structure of the bath, and not by an interplay with polarization as the case of nanophotonics \emph{chiral} quantum optics~\cite{lodahl16a,mitsch14a,sollner15a}, which paves the way of observing it for other setups beyond the optical one \cite{astafiev10a,hoi11a,vanloo13a,liu17a,devega08a,navarretebenlloch11a}. 
 
 \item We have also characterized the super and subradiant phenomena when two QEs are coupled to the bath beyond the perturbative regime. We show how collective emission can emerge, also accompanied by collective dipole-dipole interactions, and how non-perfect subradiant states are able to cancel emission in one of directions and emit mostly in the orthogonal one.
 
 \item We show how to engineer \emph{perfect} subradiant states by coupling $N=4$ QE in a square like distribution, in which the emission is trapped due to perfect destructive interference. We have also characterized them in both the Markovian and non-Markovian regimes studying the effect of retardation.
 
 \item We have studied the characteristics of superradiance when many QEs are placed along a diagonal and their energies are matched to the situation of anisotropic decay. We have seen that non-Markovian corrections provide a bound for collective emission for large $g/J$ ratios, and how collective emission can be recovered in the limit for $g/J\ll 1$.
 
 \item We have studied the impact of bath and QE loss in the observation of the phenomena described along the manuscript, providing the conditions to observe them.
\end{itemize}

Despite the simplification of the model used for the bath, we believe this paper can be used as a solid basis to understand the dynamical features of QEs interacting with real 2D photonic crystals~\cite{joannopoulos_book95a}, where the divergences associated to saddle points of $\omega(\kk)$ have also been predicted~\cite{mekis99a}.

Moreover, even though we explore many different phenomena, there are many future lines of work that one may follow, such as:
\begin{itemize}
 \item Extending the results to other geometries, e.g., triangular, and going beyond the single excitation regime or even higher dimensions.
 \item Studying the interplay between the geometry and polarization in real dielectric materials as in 1D setups \cite{lodahl16a,mitsch14a,sollner15a} to further taylor spontaneous emission.
  \item Introduce driving into the QEs and study the interplay between the anisotropic dissipation and collective driving. We foresee the existence of novel many-body 2D entangled steady states as it occurs with 1D chiral waveguides \cite{ramos14a}.
  
\end{itemize}

\acknowledgements{\emph{Acknowledgements.} The work of AGT and JIC was funded by the EU project SIQS and by the DFG within the Cluster of Excellence NIM. AGT also acknowledges support from Intra-European Marie-Curie Fellowship NanoQuIS (625955). We also acknowledge discussions with T. Shi, Y. Wu, S.-P. Yu, J. Mu\~{n}iz, and H.J. Kimble. }

\bibliography{Sci,books}

\end{document}